\documentclass[twocolumn]{aastex631}
\usepackage{amsmath}
\usepackage{enumerate}
\usepackage{enumitem}   
\usepackage{color, colortbl}
\usepackage{subfigure}
\usepackage{color}
\usepackage{courier}
\usepackage{mathtools}
\usepackage{threeparttable}
\usepackage{enumitem}
\usepackage{rotating}
\usepackage{tabularx}
\usepackage{tabularx}
\usepackage{rotating}
\usepackage{CJK}
\usepackage{multirow}
\usepackage[]{threeparttable}

\newcommand\clearrow{\global\let\rowmac\relax}

\newcommand{\Mstar}{$M_{\ast}$}
\newcommand{\OH}{O/H}
\newcommand{\dOH}{$\Delta$O/H}

\newcommand{\ppm}{\emph{p/m}}
\newcommand{\rp}{$r_\mathrm{p}$}
\newcommand{\dssfr}{$\Delta$log(sSFR)}
\newcommand{\Reff}{$R_\mathrm{e}$}

\received{}
\revised{}
\accepted{}
\submitjournal{}

\begin{document}

\title{SDSS-IV MaNGA: Spatial Evolution of Gas-Phase Metallicity Changes  Induced by Galaxy Interactions}

\correspondingauthor{Hsi-An Pan}

\begin{abstract} 
Gas-phase metallicity in interacting and merging galaxies offers key insights into their star formation processes and evolutionary histories. This study investigates the spatial evolution of gas-phase metallicity (i.e, oxygen abundance, 12 $+$ log(O/H)) in these galaxies using integral field unit (IFU) data from the SDSS-IV MaNGA survey, focusing on changes in metallicity gradients across different stages of interactions -- from early encounters to final coalescence. By comparing interacting and merging galaxies with isolated counterparts, we identify characteristic trends in how interactions influence metallicity gradients over time. Our analysis reveals that metallicity gradients typically flatten shortly after the first pericenter passage, likely due to radial gas mixing, with later stages showing either metallicity enrichment or dilution depending on the intensity of the interaction and star formation activity. These changes can result in gradients that are either flatter or steeper than the initial profiles. Notably, we observe steeper metallicity gradients in interacting galaxies at certain merger stages, which is inconsistent with predictions from some galaxy simulations. This discrepancy emphasizes the complexity of galaxy interactions.  Overall, our findings provide valuable insights into how galaxy interactions reshape metallicity distribution, enhancing our understanding of the processes driving galaxy evolution during mergers.

\end{abstract} 

\begin{CJK}{UTF8}{bkai}
	
\email{hapan@gms.tku.edu.tw}
	
\author[0000-0002-1370-6964]{Hsi-An Pan (潘璽安)}
\affiliation{Department of Physics, Tamkang University, No.151, Yingzhuan Road, Tamsui District, New Taipei City 251301, Taiwan }

\author[0000-0001-7218-7407]{Lihwai Lin}
\affiliation{Institute of Astronomy and Astrophysics, Academia Sinica, Taipei 10617, Taiwan}

\author[0000-0001-6444-9307]{Sebasti\'{a}n F. S\'{a}nchez}
\affiliation{Instituto de Astronom\'{i}a, Universidad Nacional Aut\'{o}noma de M\'{e}xico, A.P. 106, Ensenada, 22800 BC, M\'{e}xico}

\author[0000-0003-2405-7258]{Jorge K. Barrera-Ballesteros}
\affiliation{Instituto de Astronom\'{i}a, Universidad Nacional Aut\'{o}noma de M\'{e}xico, A.P. 70-264, 04510 M\'{e}xico, D.F., M\'{e}xico}

\author[0000-0001-5615-4904]{Bau-Ching Hsieh}
\affiliation{Institute of Astronomy and Astrophysics, Academia Sinica, Taipei 10617, Taiwan}

\section{Introduction}
Understanding the evolution of galaxies involves numerous complex processes, including the distribution and enrichment of metals -- elements heavier than helium -- in the interstellar medium (ISM). 
In the inside-out formation model \citep{Pee69,Lar76}, the most intense and earliest star formation typically occurs in the central regions of galaxies. 
This star formation enriches the ISM with metals produced by stellar nucleosynthesis in these central areas, resulting in gas-phase metallicity that is highest at the center and decreases with increasing radius,  exhibiting a negative gradient \citep[e.g.,][]{San14,Bel17,San20,Boa23}.
As galaxies evolve, their gas-phase metallicity undergoes significant changes by star formation and feedback mechanisms \citep{Boa24,Wan21,Sha24},  and gas inflow and outflow processes \citep{Sha21a,Bas24}.
These changes in metallicity and its distribution thus offer crucial insights into the processes that drive galaxy evolution.

Mergers play a crucial role in the life cycle of galaxies. 
Galaxy formation through hierarchical clustering.
In this process,  smaller structures merge over time to form increasingly larger and more complex galaxies.
Observationally, around 10 -- 50\% of galaxies are interacting with other galaxies \citep{Pat02,Lin04,Lin08}. 
The merger fraction depends on redshift, stellar mass, and the definition of galaxy mergers \citep[e.g.,][]{Kim21,Nev23}.
During these dynamic interactions, galaxies can experience significant transformations, including bursts of star formation, morphological changes, and redistribution of gas and metals \citep{Too77,Dim07,Loz08a,Loz08b}. 
Therefore, observing and modeling the spatial evolution of gas-phase metallicity across various merger stages can enhance our understanding of how these common interactions influence galactic properties over time.

Previous studies based on integrated galaxy properties have shown that the centers of interacting galaxies tend to be metal-poor compared to galaxies of similar mass \citep{Kew06, Scu12}. Theoretically, during a merger, low-metallicity gas from the outer regions is funneled into the high-metallicity central regions due to gravitational torques, leading to a lower average abundance in the core. This redistribution of gas alters the metallicity profile across the entire disk, effectively washing out the original metallicity gradient.
However, the detailed spatial evolution of metallicity and its dependence on the merger stage remain poorly understood observationally due to limited measurements in the outer disk regions.
Simulations predict that the radial metallicity gradients of disk galaxies flatten shortly after each passage \citep{Rup10,Mon10,Per11,Por22}. 
This flattening reflects the impact of gas redistribution across the galactic disk.
However, interaction-triggered star formation \citep{Ell08}, which enriches the ISM, adds further complexity to the radial metallicity gradients. 
The subsequent metallicity gradients therefore depend on the balance between radial mixing and ISM enrichment at different galactic radii.

The flattening of metallicity gradients has indeed been observed by \cite{Kew10} using \ion{H}{2} region metallicities obtained from high signal-to-noise multi-slit observations. 
However, the results of \cite{Kew10} are limited by the sample size and thus the range of merger stages they can cover. 
Moreover, observing individual \ion{H}{2} regions restricts the coverage of the galactic disk, providing data only for the observed regions. 
The selection of \ion{H}{2} regions might also introduce bias, leading to an incomplete understanding of the overall metallicity distribution. 
Consequently, there is still a lack of a well quantified picture  of merger-driven  evolution of metal distribution from observations, which are important to constrain the theoretical models of metal enrichment, feedback prescription, and galaxy evolution.

By examining galaxies at various stages of merging, from early encounters to coalescence, we aim to construct a comprehensive picture of how gas-phase metallicity evolves spatially.
In this work, the gas-phase metallicity is indicated by the abundance of oxygen (O) relative to hydrogen (H), typically expressed on a logarithmic scale as 12~$+$~$\log$(O/H) (hereafter O/H).
The metallicity measurements are taken using integral field unit (IFU) observations from the SDSS-IV MaNGA survey \citep{Bun15}.
The IFU technique provides spatially-resolved maps of metallicity across most of the galactic disk and simultaneously collects data from numerous regions, significantly reducing the overall observation time.
Consequently, it increases the sample size and the range of merger stages that can be studied.
\citet{Bar15} examines the impact of galaxy interactions on  oxygen abundance using IFU data from the CALIFA survey, focusing on the central regions. They find similar oxygen abundances compared to non-interacting galaxies. The study focuses only on the central regions rather than the spatial distribution across galaxy disks.
In this work, we will analyze the evolution of metallicity gradients in interacting galaxies and compare them to isolated counterparts to identify and quantify the characteristic trends.

The structure of the paper is as follows. Section \ref{sec_data} presents the data and analysis, including the datasets used for metallicity measurements, merger stage classification, and quantification of interaction-induced metallicity changes.
The results are presented and discussed in Section \ref{sec_result}. 
The main findings are summarized in Section \ref{sec_summary}.
Throughout this paper, we adopt the following cosmological parameters: $H_\mathrm{0}$ = 70 km s$^{-1}$ Mpc$^{-1}$, $\Omega_\mathrm{m}$ = 0.3, and $\Omega_\mathrm{\Lambda}$ = 0.7.

\section{Data and Analysis}
\label{sec_data}
In this work, we will study the spatial evolution of interaction-induced metallicity changes in interacting galaxies.
The data and analysis methods follow our previous work on the spatial evolution of interaction-triggered star formation in \cite{Pan19}.
Here we briefly summarize the data used in this work and the classification of merger stages.

\subsection{MaNGA Observation and Metallicity Measurement}
We employ data from SDSS-IV MaNGA  (Mapping Nearby Galaxies at Apache Point Observatory; \citealt{Bun15}) to explore gas-phase metallicity in interacting galaxies.
The MaNGA data provides detailed spatially resolved spectral information by its integral field spectroscopy. 
This allows for the precise measurement of gas-phase metallicity at different locations within each galaxy.   
The spectral resolutions vary from  R $\sim$ 1400 at 4000\,\AA\, to R$\sim$ 2600 around 9000\,\AA.
The point spread function (PSF)  is $\sim$ 2.5$\arcsec$, corresponding to 1.8 kpc at the median redshift of the MaNGA sample ($z$~$\sim$~0.03).
For more details on the MaNGA setup, we refer the reader to \citet{Dro15} for the IFU fiber feed system, \citet{Wak17} for the sample selection, \citet{Law15} for the observing strategy, and \citet{Law16}, \citet{Wes19}, and \citet{Bel19} for the MaNGA data reduction and analysis pipelines. We selected galaxies from a sample of 4,691 galaxies observed by MaNGA within the first four years of operation, corresponding to the SDSS Data Release 15 \citep{Agu18}.
The reduced MaNGA datacubes are analyzed using the Pipe3D pipeline to extract the physical parameters from each spaxel of each galaxy \citep{San16a, San16b, San18}. 
Each spaxel covers an area of 0.5$\arcsec$ $\times$ 0.5$\arcsec$. Pipe3D fits the continuum with stellar population models and measures the nebular emission lines. 

A commonly used proxy for  gas-phase metallicity is the oxygen abundance, measured from emission lines in star-forming regions of galaxies. As oxygen is the most abundant heavy element, its abundance serves as a reliable estimate for the overall metal content.
The gas-phase metallicity in this work is determined using the \emph{O3N2} calibration by \cite{Mar13}, ensuring consistency with previous studies on metallicity gradients in the MaNGA sample by \cite{Bel17} and \cite{Boa23}, as well as with the metallicity gradients in MaNGA post-mergers studied by \cite{Tho19}\footnote{The \emph{O3N2} is defined as $\textit{O3N2} = \frac{[\text{O\,\textsc{iii}}]\lambda5007/\text{H}\beta}{[\text{N\,\textsc{ii}}]\lambda6584/\text{H}\alpha}$. However, it is important to note that while \cite{Bel17} use the \emph{O3N2} calibration from \cite{Pet04}, \cite{Tho19}, \cite{Boa23}, and this work employ the \emph{O3N2} calibration from \cite{Mar13}. The use of the O3N2 calibrator by \cite{Pet04} tends to yield larger gradients compared to more recent calibrators, such as \cite{Mar13}. 
}.
The \emph{O3N2} index is advantageous because it varies monotonically with metallicity and uses line ratios that are close in wavelength, reducing uncertainty from extinction correction or flux calibration errors \citep{Mar13}.
Numerous studies have compared metallicity derived from different calibrators.
While there may be systematic offsets in metallicity between different calibrators, the various methods generally agree qualitatively \citep{Kew06,Kew08,Poe18,Kew19,Scu21, Tei21}. 
Since our main results will be based on the offset of metallicity relative to control isolated galaxies, the exact value of metallicity is not the primary concern. 
Nonetheless, we will demonstrate in the Appendix \ref{appendix_oh_cal} that our main results can be reproduced even when different calibrators are used.

Since metallicity calculations are typically applied only to star-forming regions, we use only star-forming spaxels in the analysis of this work. Star-forming spaxels are selected using excitation diagnostic diagrams \citep{Kew01,Kau03,Cid13} and an H$\alpha$ equivalent width cut of $>$ 6~\AA\ \citep{San14}. Additionally, we restrict the analysis to spaxels with a signal-to-noise ratio (S/N) greater than 3 for the emission lines required to compute metallicity using the \emph{O3N2}, \emph{N2}, and \emph{N2S2} methods. The latter two methods are used to ensure that our main results remain robust regardless of the choice of metallicity calibrator (see Appendix \ref{appendix_oh_cal}).
Since we further limit our analysis to star-forming galaxies (see next section), more than 90\% of the spaxels in our sample are classified as star-forming.

\subsection{Merger Sequence}
\begin{figure*}
	\centering
	\includegraphics[scale=0.3]{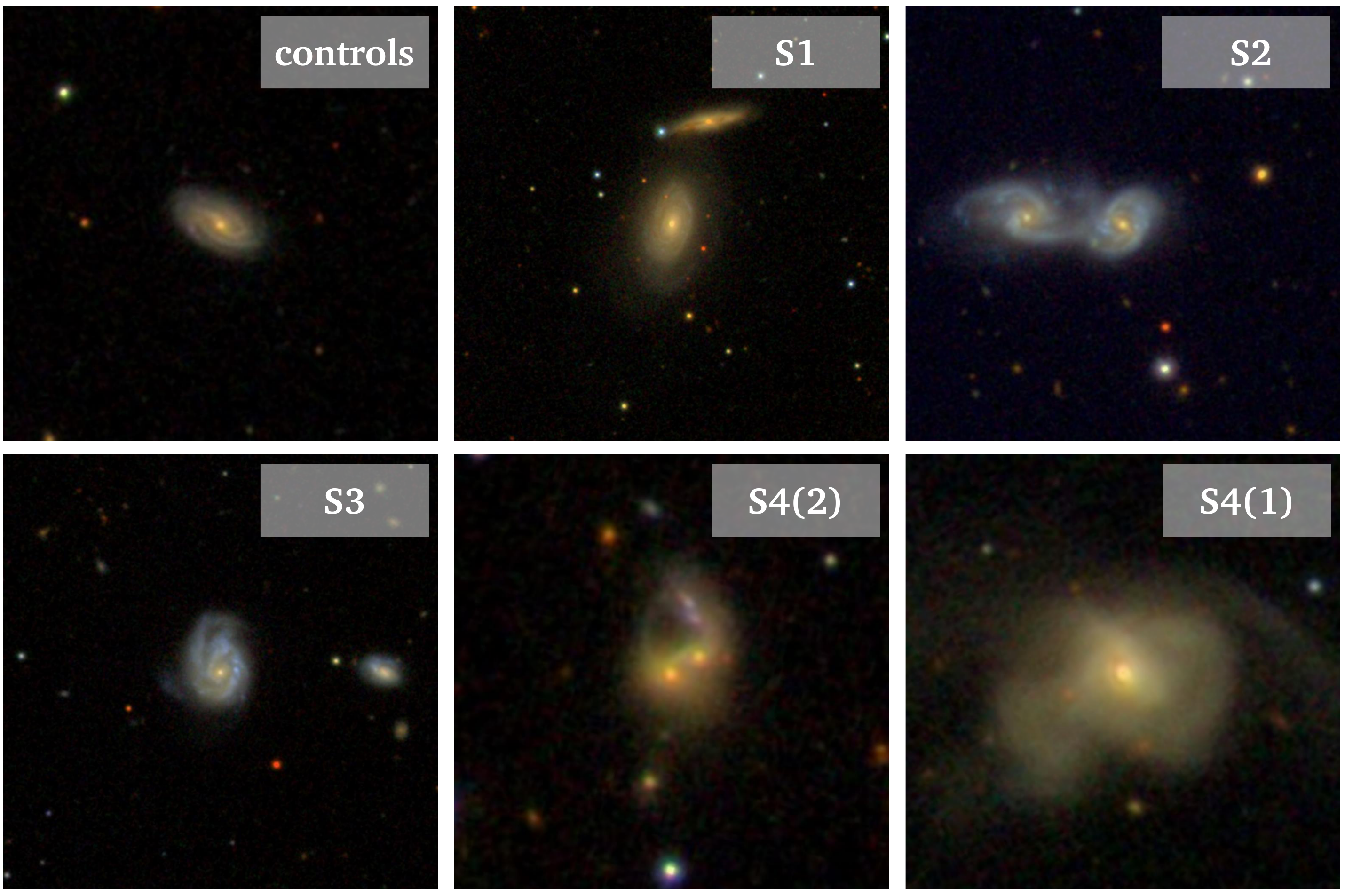}
	\caption{Merger Stage Classification: Stage 1 comprises galaxies in well-separated pairs with no morphological distortion. Stage 2 includes close pairs exhibiting strong interaction features, such as tails and bridges, indication the first pericenter passage. Stage 3 consists of well-separated pairs with mild morphological distortion, likely near or just past the apocenter. Stage 4 features strongly overlapping components with significant morphological distortion. Within Stage 4, Stage 4(2) refers to cases with two visible nuclei, while Stage 4(1) includes cases with only one visible nucleus, representing post-mergers. Finally, the control sample consists of galaxies with no spectroscopic companions and symmetric disks, as revealed by SDSS \emph{gri} composite images. This classification follows the same criteria as in our previous paper \citep{Pan19}.
	}  
	\label{fig_merger_stages}
\end{figure*}

We determine the merger stages through visual inspections of the $gri$ composite images observed by the 2.5-m Telescope of SDSS.
Examples of each stage are presented in Figure \ref{fig_merger_stages}. 
Further discussion on the design of this classification scheme can be found in our previous work \citep{Pan19}.
In this work, interactions between galaxies are classified according to the following scheme:

\begin{itemize}
	
	\item {\bf Stage 1} -- well-separated pair  which do not show any morphology distortion (i.e., incoming pairs, before the first pericenter passage),
	
	\item {\bf Stage 2} -- close pairs showing strong signs of interaction, such as tails and bridges (i.e., at the first pericenter passage),
	
	\item {\bf Stage 3} -- well-separated pairs, but showing weak morphology distortion (i.e.,  approaching the apocenter or just passing the apocenter),
	
	\item {\bf Stage 4} --  two components strongly overlapping with each other and show strong morphological distortion (i.e., final coalescence phase), or single galaxies with obvious tidal features such as tails and shells (post-mergers),
	
\end{itemize}

\subsection{Sample}
The selection of pair and control samples is identical to that in \cite{Pan19}. 
The selection scheme is summarized here.

\subsubsection{Galaxies in Pairs and Mergers}
\label{sec_sample_pm}

Galaxies in pairs or mergers (\emph{p/m}) are defined as those with a spectroscopic companion. Galaxies in pairs must have a companion at a projected separation of $<$50 kpc h$^{-1}$ (or 71.4 kpc) and a line-of-sight velocity difference of $<$500 km s$^{-1}$ \citep[e.g.,][]{Pat02,Lin04}. However, if the two components are too close (i.e., Stage 2 and Stage 4) to be de-blended by SDSS for separate spectroscopic redshifts, or if they are post-mergers (i.e., single galaxies), they are missed by the spectroscopic criteria. Therefore, we visually inspect the SDSS \emph{gri} composite images of all MaNGA galaxies to recover these closest interacting systems.

Galaxies that are quenching or have already quenched are excluded from this study. 
We apply a criterion of global specific star formation rate of log(sSFR/yr$^{-1}$) $>$ $-$11 to select star-forming galaxies, where sSFR is defined as the global  star formation rate (SFR) divided by the stellar mass (\Mstar). 
It is important to note that we do not require the companion galaxies in our sample to be star-forming.
The sample for this paper comprises 205 galaxies in \emph{p/m}.
The fraction (and number) of galaxies in Stage 1 through Stage 4 are 19\% (38), 12\% (24), 28\% (58), and 41\% (85), respectively. Of the galaxies in Stage 4, 15\% (13) have visually double nuclei, while 85\% (72) have a single nucleus. 
Hereafter, we refer to Stages 1 through 4 as \textit{S1} through \textit{S4}, respectively. 
For \textit{S4}, we use \textit{S4(2)} and \textit{S4(1)} to represent the sub-categories with double nuclei and a single nucleus, respectively.

\subsubsection{Control Sample}
\label{sec_sample_control}

We define a control sample to quantify the effect of interactions.
To construct a reliable control sample, we not only use spectroscopic data but also the "P-merger" parameter from Galaxy Zoo to exclude potential interacting galaxies based on their morphology \citep{Dar10a,Dar10b}. 
The P-merger parameter quantifies the probability that an object is a merger, ranging from 0 (an object that does not resemble a merger) to 1 (an object that unmistakably looks like a merger).
Control galaxies should have no spectroscopic companion,  a P-merger value of 0, \emph{and} a global specific star formation rate log(sSFR/yr$^{-1}$) $>$ $-$11. 
A total of 1348 MaNGA star-forming galaxies are selected as control galaxies.

\subsection{Characteristics of metallicity profiles of the controls}
\label{sec_chara_controls}

We first examine the metallicity profiles in the control group (i.e., isolated galaxies) to ensure that our derived profiles accurately capture the known characteristics and therefore serve as a reasonable reference.

\begin{figure*}
	\begin{center}
		\subfigure[]{\label{fig_rad_vs_OH_ctrls}\includegraphics[width=0.49\textwidth]{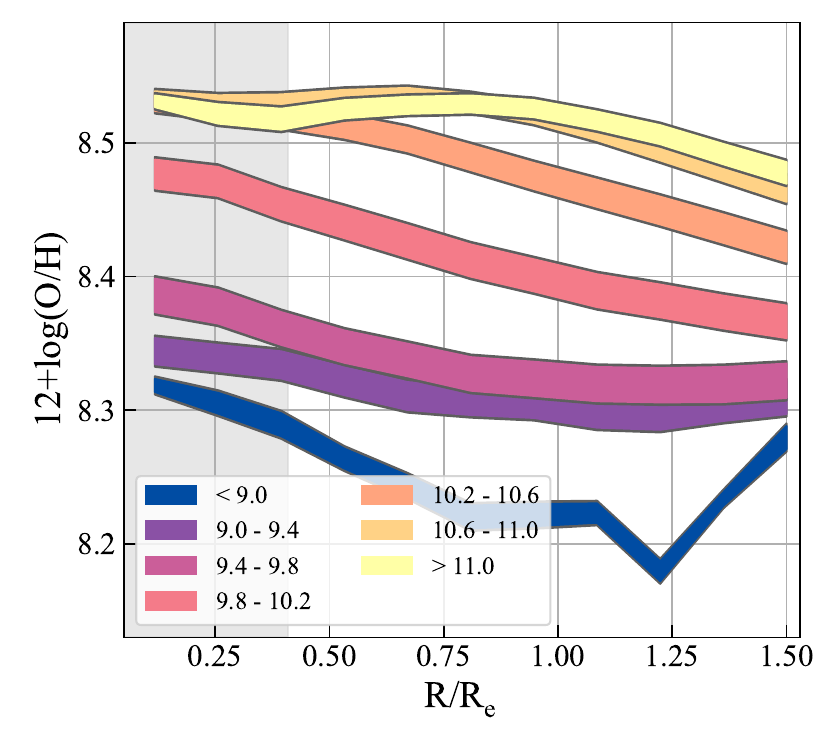}}
		\subfigure[]{\label{fig_Mstar_OH_slope}\includegraphics[width=0.49\textwidth]{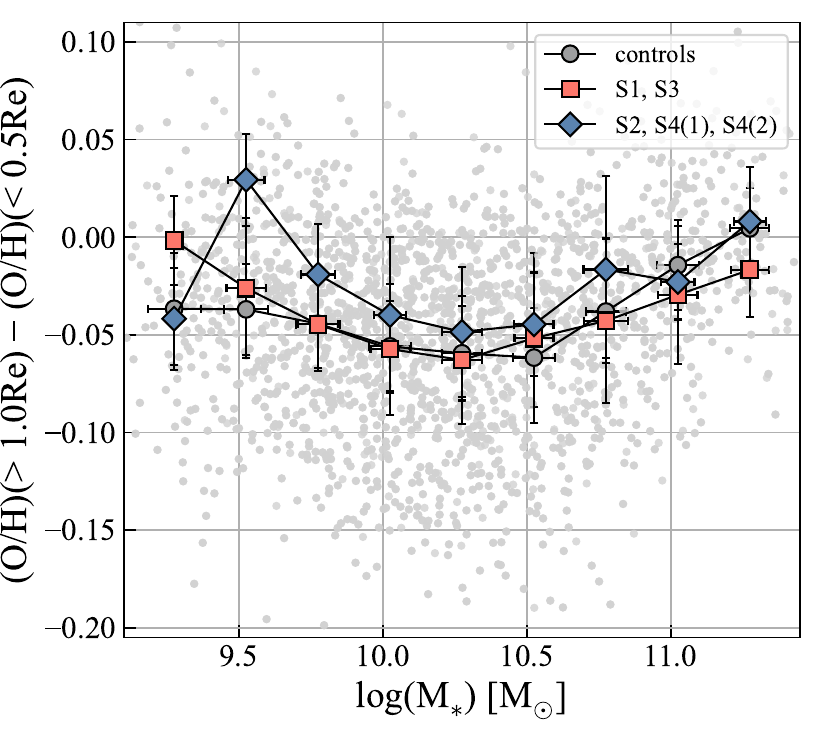}}
	\end{center}
	\caption{\OH\ characteristics of the control galaxies. 
		(a) Radial profiles of oxygen abundance  as a function of radius  for galaxies in different \Mstar\ bins. 
		The gray shaded area represents the 2.5$\arcsec$ FWHM of MaNGA observations, corresponding to $\sim$0.4~\Reff\ of our sample. 
		The color-coded lines represent different stellar mass ranges, as indicated in the legend. The color shaded regions correspond to the uncertainty, error of the mean, in the measurements.  Galaxies with $\log$(\Mstar/M$_{\sun}$) $<$ 10 and $\log$(\Mstar/M$_{\sun}$) $>$ 10.5 exhibit shallower profiles compared to those with intermediate \Mstar.
		(b) Comparison of the \OH\ gradients as a function of \Mstar\ across different sample groups. The \emph{gradient} is defined as the difference in \OH\ between the outer region ($>$ 1.0 \Reff) and the inner region ($<$ 0.5 \Reff) of the galaxies. Control galaxies are represented by large gray circles, while Stage 1 and Stage 3 galaxies are shown as red squares, and Stage 2, Stage 4(1), and Stage 4(2) galaxies are depicted as blue diamonds. The error bars show the standard error of the mean. The gray dots in the background represent individual control galaxies.	}
	\label{fig_controls_oh}
\end{figure*}

Figure \ref{fig_rad_vs_OH_ctrls} illustrates the radial \OH\ profiles for the control galaxies across various \Mstar\ groups. 
The figure shows the rank of local \OH\ with \Mstar\ across nearly all radial extents under consideration. 
Additionally, diverse \OH\ profiles are evident among galaxies of different \Mstar. 
Galaxies with $\log$(\Mstar/M$_{\sun}$) $<$ 10 and $\log$(\Mstar/M$_{\sun}$) $>$ 10.5 exhibit shallower profiles, while those with intermediate \Mstar\ display steeper profiles.
The gray shaded area in Figure \ref{fig_rad_vs_OH_ctrls} represents the 2.5$\arcsec$ FWHM of the MaNGA observations, corresponding to $\sim$ 0.4~\Reff\  of our sample. While the radial profiles are resolved overall, the profile within 0.4~\Reff\ should be interpreted with caution.

We calculate the \emph{slope} of \OH\ by comparing the median \OH\ in the inner region ($<$ 0.5~\Reff) with that in the outer disk region (1.0 -- 1.5~\Reff). The choice of $<$ 0.5~\Reff\ captures the central region, typically dominated by the bulge in galaxies (as well as the unresolved areas of the MaNGA observation), where star formation and metallicity distributions may differ from those in the outer disk. By comparing this with the 1.0 -- 1.5~\Reff\ range, we can contrast the central and outer regions without needing to fit a gradient, which avoids the dependence on the specific mathematical form used for the fitting.
The derived \emph{slope} as a function of \Mstar\ is shown in Figure \ref{fig_Mstar_OH_slope}, where small gray circles represent individual control galaxies, and larger gray circles denote the median \OH (1.0~$<$~$R$~$<$~1.5~\Reff)  $-$ \OH ($<$~0.5~\Reff)  values for different \Mstar\ bins. 
Figure \ref{fig_Mstar_OH_slope} shows a mild curvature around $\log$(\Mstar/$\mathrm{M_{\sun}}$) $=$ 10.0 -- 10.5, with flat gradients on either side, mirroring the trend observed in Figure \ref{fig_rad_vs_OH_ctrls}.
This pattern aligns well with previous analyses of MaNGA data \citep{Bel17,Min20,Boa23} and other IFU surveys like CALIFA and SAMI \citep{San14,Zin19}, also in agreement with model prediction \citep{Sha21b}. This consistency confirms that the selected control galaxies are well-suited to serve as a reliable reference for isolated galaxies.

Nonetheless, we should note that the moderate spatial resolution of MaNGA observations, along with the presence of Diffuse Ionized Gas (DIG), could potentially influence the derived metallicity values and, consequently, the resulting metallicity gradient.
Our derived metallicity profiles and slopes for MaNGA galaxies may be influenced by the moderate spatial resolution of 1 kpc, particularly in smaller galaxies or more compact late-stage mergers. Limited resolution can affect the ability to fully resolve individual HII regions, potentially leading to some smoothing or flattening of metallicity gradients \citep{Ach20}.
However, simulations of galaxy observations across different redshifts (and corresponding spatial resolutions) indicate that kpc-scale resolution remains effective in capturing key observables in galaxy studies \citep{Mas14}. While minor smoothing of the gradient slope may occur, the overall structure and trends remain reliable.

A valuable test of resolution effects could be conducted using high-resolution ($\sim$ 70~pc) MUSE IFU data from the PHANGS survey \citep{Ems22}. By systematically degrading the resolution and analyzing its impact on metallicity profiles, we can directly assess how spatial averaging influences observed gradients. Such an approach would provide important insights into the robustness of metallicity trends and help refine interpretations of lower-resolution datasets like MaNGA.

DIG, which has weaker emission line fluxes compared to star-forming regions, can impact the accuracy of metallicity measurements by diluting the observed emission, particularly in the outer regions of galaxies where DIG contributes more significantly. This contamination can either steepen or flatten metallicity gradients, depending on the galaxy and the specific metallicity caolibrator used.
The effect of DIG on \OH\ measurements varies based on the strong-line index applied. Among commonly used indices for estimating oxygen abundance, the \emph{N2} index is most affected by DIG contamination, leading to a potential bias of up to 0.1 dex in O/H. In contrast, the \emph{O3N2} index, which we use in this study, is less sensitive to DIG effects \citep{Kum19,Val19}. This is because both [O,\textsc{iii}]$\lambda$5007/H$\beta$ and [N,\textsc{ii}]$\lambda$6584/H$\alpha$ tend to rise in tandem in regions dominated by DIG, reducing the impact on the derived metallicity.
As a result, while DIG contamination is a known factor in metallicity measurements, our use of the \emph{O3N2} calibrator ensures that the effect is minimal compared to other calibrators.

\subsection{Metallicity Offset (\dOH)}
\label{sec_offset_deine}
To quantify the interaction-induced metallicity change, we compute the metallicity ``offset'' (\dOH) for each spaxels. 
The method generally follows the approach  in our previous work \citep{Pan19} for analyzing spatial evolution of interaction-triggered star formation, as well as other studies that utilize this ``offset'' metric for either star formation \cite[e.g.,][]{Ell18} or metallicity analysis \cite[e.g.,][]{Tho19}.
For a given spaxel, control spaxels from the control galaxies are identified based on several global and local criteria: they must match within 0.005 in redshift, 0.1 dex in global \Mstar, 0.1 dex in local $\Sigma_\ast$, 20\% in galaxy size, and 0.1 in spaxel position, with both dimensions expressed in units of effective radius (\Reff, the radius which contains half the total light in $r$-band of the galaxy).
The rationale for this approach, rather than relying solely on local properties given our focus on spatially-resolved characteristics, is discussed in the Appendix \ref{appendix_matching} (see also \citealt{Ell18} and \citealt{Tho19}).

Once the control spaxels are established, the median \OH\ value of these spaxels is subtracted from the value of the spaxel in question to calculate \dOH\ as follows:
\begin{equation}
\Delta \text{O/H} = \text{O/H} -  \text{median}(\text{O/H}_{\text{controls}}).
\end{equation}
The \dOH\ value is also calculated for each spaxel in the control galaxies to validate the matching scheme.
Since \OH\ is in logarithmic form, a positive \dOH\ represents an enhancement of \OH\ with respect to the controls, and vice versa.

Finally, the radial \dOH\ profile for each galaxy is computed with radial bins of 0.15\Reff. 
The radial \dOH\ extends out to 1.5\Reff, matching the radial coverage of the MaNGA fiber bundles, which is at least 1.5\Reff. 
We calculate the radial \dOH\ distribution for both \ppm\ and control samples.



\section{Results and Discussion}
\label{sec_result}
In this section, we will first present the statistical results derived from spaxels before turning into the findings concerning radial metallicity profiles.

\subsection{Spaxel-by-Spaxel distributions}
\label{sec_spaxel_distribution}

\begin{figure*}
	\begin{center}

		\subfigure[]{\label{fig_boxplots}\includegraphics[width=0.96\textwidth]{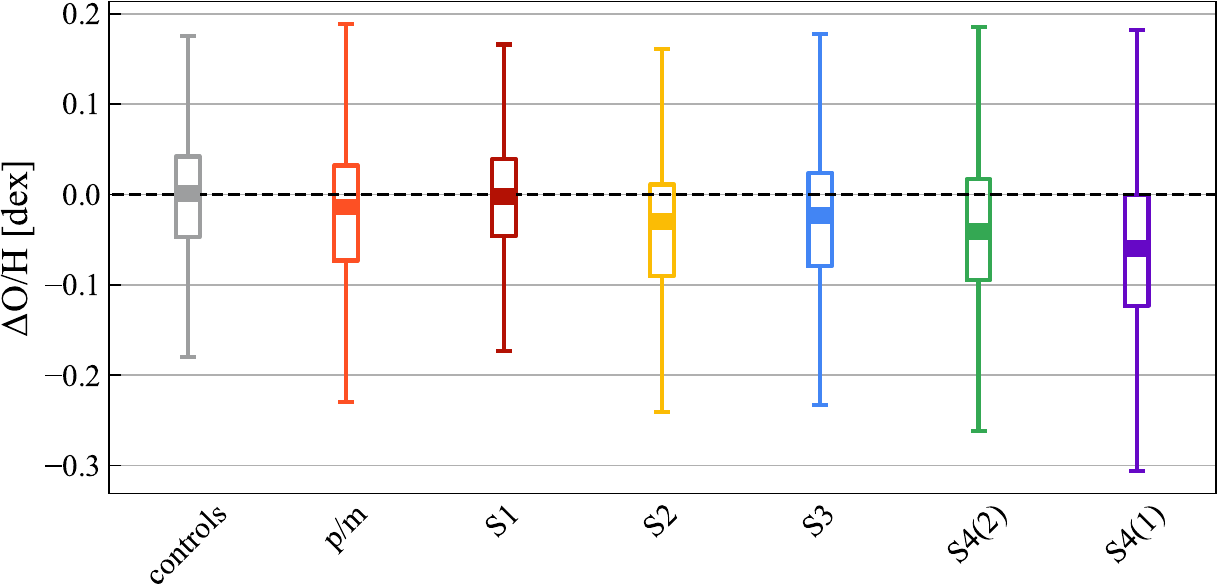}}
		\subfigure[]{\label{fig_boxplots2}\includegraphics[width=0.96\textwidth]{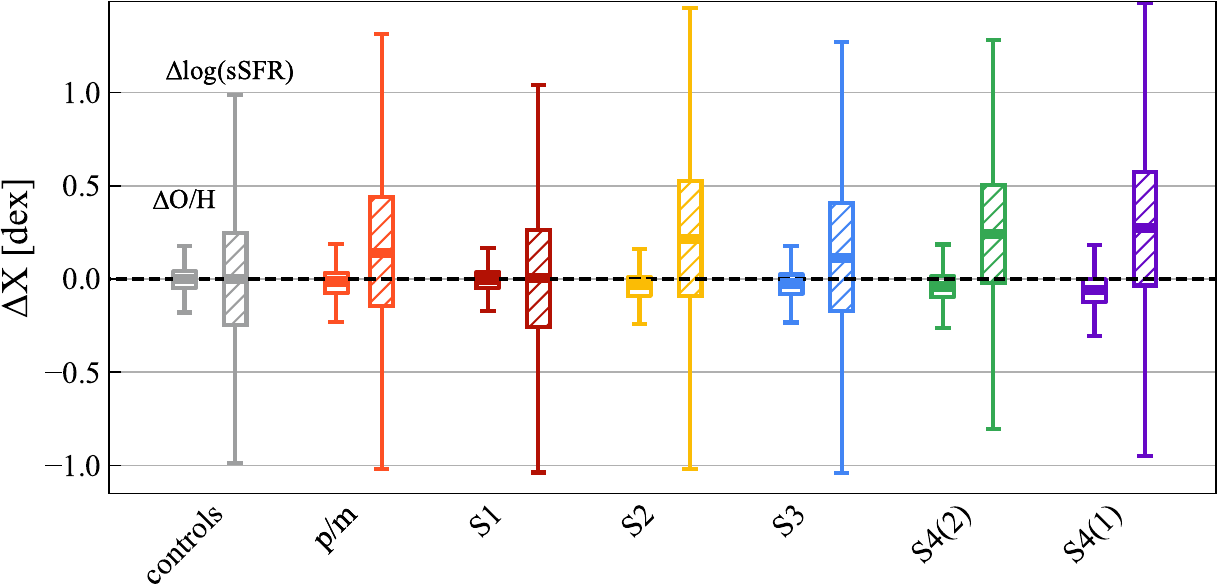}}
	\end{center}
	\caption{
	(a) Boxplot of \dOH\ differences across various galaxy groups: controls, all \ppm, and merger stages S1 through S4(1).  The dashed line at \dOH~$=$~0 indicates no change in metallicity. In each boxplot, the median is represented by the solid square at the center. The ends of the box mark the upper and lower quartiles, encompassing the interquartile range, which contains 50\% of the sample. The vertical whiskers  extend beyond the box to 1.5 $\times$ IQR. The figure demonstrates how metallicity changes over time during galaxy interactions.
	(b) Boxplot comparison of \dOH\ (solid boxes) and \dssfr\ (hatched boxes) across different galaxy groups. This anti-correlation in median values is consistent across the various groups, suggesting a potential underlying physical mechanism that links metallicity and star formation processes during galaxy interactions. Note that the $y$-axis ranges differ between the two panels.
	}
	\label{fig_main}
\end{figure*}

Figure \ref{fig_boxplots} illustrates the spaxel-based \dOH\ distribution across different merger stages using box plots. 
In each boxplot, the median is indicated by the solid squares in the middle. 
The ends of the box are the upper and lower quartiles (the interquartile range, IQR); 50\% of the sample is located inside the box. 
The two whiskers (vertical lines) outside the box extend to 1.5$\times$IQR.
Boxplots from left to right represent the distribution for the
galaxies in the control pool, all the galaxies in \ppm\, and S1 -- S4(2), respectively.

The distribution of the control group centers around zero, with a median of $+$0.002~dex, verifying the accuracy of the \dOH\ calculations. 
In contrast, the median \dOH\ for galaxies in the \ppm\ sample is $-$0.021 dex, which is lower than that of the controls. 
The S1 through S4(1) groups show varying distributions, with the median value changing across the merger stages. 
Specifically, the median \dOH\ decreases from $+$0.006 for S1 to $-$0.040 for S2, then slightly increases to $-$0.033 for S3, drops to $-$0.060 for S4(2), and further decreases to $-$0.086 for S4(1). Notably, in the case of S4(1), the entire box is below zero, indicating a widespread reduction in \OH\ for post-merger systems. Figure \ref{fig_boxplots} clearly demonstrates that galaxies in \ppm\ generally exhibit lower metallicity compared to control isolated galaxies, with a reduction in metallicity observed at all stages except for S1, though with varying magnitudes.

Figure \ref{fig_boxplots2} presents boxplots comparing the distribution of \dOH\ across different sample groups, along with the corresponding \dssfr\ \citep{Pan19} distributions indicated by the hatched boxplots\footnote{The \dssfr\ is calculated in the same way as described in Section \ref{sec_offset_deine}. We refer the reader to our previous paper \citep{Pan19} for the details.}. 
The figure reveals a clear anti-correlation between \dOH\ and \dssfr, where higher values of \dOH\ tend to coincide with lower \dssfr, and vice versa. 
Specifically, the observed trend implies that regions or galaxies with enriched oxygen levels tend to exhibit lower sSFR, while those with depleted oxygen levels show enhanced star formation activity. 
This anti-correlation is consistent across the various  groups, highlighting a potential underlying physical mechanism linking metallicity and star formation processes during galaxy interactions.

\subsection{Spatial Distribution of \dOH}
\label{sec_spatial_distribution}
\subsubsection{Slopes}
In galaxy simulations, interactions lead to the radial redistribution of gas, resulting in a reduction of radial \OH\ gradients \citep[e.g.,][]{Rup10,Kar22,Tis22}. 
The inflow rate typically increases during or shortly after each close passage \citep{Tor12}. 
Therefore, we first compare the slopes, as defined in Section \ref{sec_chara_controls}, between galaxies undergoing a close passage (S2, S4(2), and S4(1)) and those in widely-separated phases (S1 and S3).

The orange squares in Figure \ref{fig_Mstar_OH_slope} display \OH($\mathrm{<0.5R{e}}$$)-$\OH$(\mathrm{0.5-1.5R{e}}$)  values against \Mstar\ for galaxies at S1 and S3, while the blue diamonds represent galaxies near first or final passage (S2, S4(2), and S4(1)).
For the S1 and S3 stages, the median \OH\ slopes are consistent with those of the control across nearly all \Mstar\ bins, with the exception of the highest and lowest regimes where the smaller sample size leads to greater uncertainties. 
Consequently, we observed no significant differences in the \OH\ slopes between galaxies pre- or post-first passage and their controls.

Conversely, galaxies undergoing close passage (S2, S4(2), and S4(1)) generally display flatter \OH\ slopes compared to control groups across most \Mstar\ bins, which is consistent with predictions from galaxy simulations \citep[e.g.,][]{Rup10}. 
Although the differences observed are not statistically significant due to the associated uncertainties, the result align with expected trends and highlight the potential influence of interactions. 
Enhancing the sample size in future studies could enable more definitive conclusions and further validate these  findings.

Figure \ref{fig_Mstar_OH_slope}  shows that the the metallicity gradient indeed being flatten for galaxies in pairs and mergers, but such flattening only occurs in  galaxies undergoing close passage (S2, S4(2), and S4(1)).
\cite{Rup10} employed N-body/smoothed-particle hydrodynamics numerical simulations to study the evolution of metallicity gradients in interacting systems. 
Their simulations revealed that the gradient typically becomes shallow or flattens immediately after the first pericenter passage, and remains flat until the final coalescence phase. 
In contrast, our observations indicate that the metallicity gradient becomes steeper, reverting closer to its initial value, after the first pericenter (S3). 
This finding contradicts the results from the simulations by \cite{Rup10}.

It is widely recognized that galaxy interactions trigger inflows of gas, which in turn drive enhanced star formation across the galaxy, peaking in the nuclear region \citep{Pan19}.
Star formation enriches the ISM through nucleosynthesis, which can potentially alter the metallicity gradient depending on where the star formation occurs. 
However, the simulations by \cite{Rup10} do not account for chemical enrichment from star formation. 
Therefore, the metallicity following a close passage (e.g., S3) might be underestimated. 
Indeed, simulations that account for chemical enrichment by star formation demonstrate that the nuclear ISM is enriched shortly (within a few hundred million years) after the first passage \citep{Mon10,Per11,Tor12}.
The \OH\ at the disk regions also vary as interaction proceeds by either enhanced in situ star formation activity and/or the metal-rich material coming from inner regions as the galactic disk get distorted \citep{Per11}.

In summary, our findings challenge the prevailing assumption that galaxy interactions uniformly flatten gas-phase metallicity profiles. Instead, we observe a more dynamic evolution: following a close passage (e.g., stage S2), the metallicity gradient initially flattens due to gas inflows and dilution, but later, in the post-interaction phase (stage S3), it can revert closer to its initial state or even steepen. This highlights the intricate interplay between galaxy interactions, star formation, feedback, and chemical enrichment.

This evolutionary cycle is well-supported by the latest cosmological simulations from the CIELO project \citep{Tis25,Tap25}, which incorporate a multiphase gas model, metal-dependent cooling, star formation, and stellar feedback. These simulations show that interaction-driven gas inflows first dilute the central metallicity, but subsequently, triggered star formation enriches the gas, steepening the metallicity gradient; at later stages, the combined effects of starburst-driven outflows and interaction-triggered inflows redistribute metals, resulting in a secondary phase of gradient flattening (see Figure 10 of \citealt{Tap25} for an example, and also results from other cosmological simulations, such as \citealt{Tor19} and \citealt{Tis22}). This sequence of events closely mirrors our observational data, reinforcing the agreement between real galaxy interactions and the latest theoretical models.


\subsubsection{Radial Profiles}
\begin{figure*}
	\centering
\subfigure[]{\label{fig_rad_dOH}\includegraphics[width=0.49\textwidth]{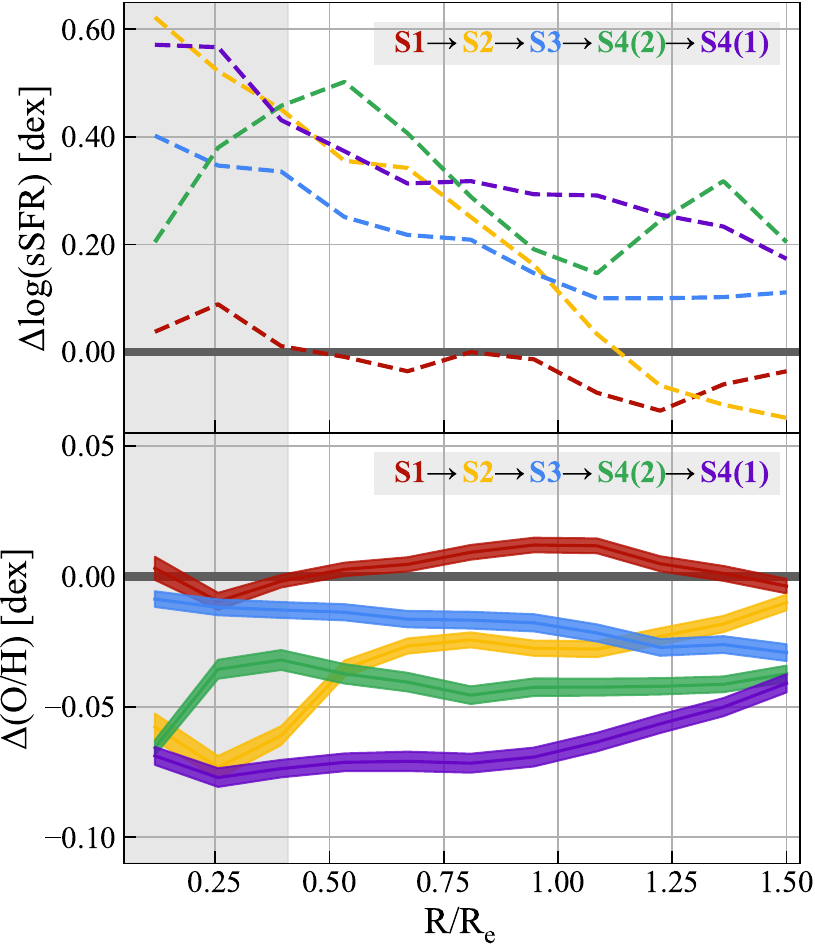}}
\subfigure[]{\label{fig_OH_vs_SFR}\includegraphics[width=0.49\textwidth]{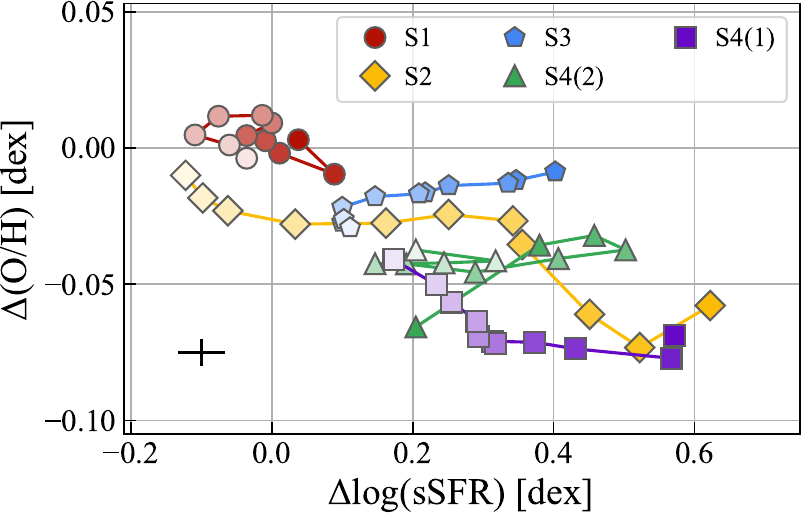}}
	\caption{Radial profiles of changes in \dssfr\ (top panel; taken from \citealt{Pan19}) and \dOH\ (bottom panel) for galaxies at different merger stages: S1 (red), S2 (orange), S3 (yellow), S4(2) (green), and S4(1) (purple).  The black horizontal line indicates the zero point. Metallicity and star formation radial profiles generally exhibit an anti-correlation, where an increase in one is typically followed by a decrease in the other.
	Moreover, \dssfr\  generally decreases with radius, while \dOH\ also varies with radius, showing different trends at various stages of the merger. The radial metallicity profile can be either negative (e.g., S2) or positive (e.g., S3) relative to the control galaxies. Please note that the $y$-axis ranges differ between the two panels. (b) \dssfr\ versus \dOH. The data points represent radial bins in panel (a), with darker points indicating regions closer to the galactic center. The typical uncertainties for \dssfr\ and \dOH\ are shown in the lower-left corner. Diverse patterns are observed between \dssfr\ and \dOH\ across different merger stages, emphasizing that galaxy interactions and mergers can induce significant, stage-dependent variations in the mass-metallicity and star formation relationship, highlighting the dynamic nature of galaxy evolution.
	}
	\label{fig_rad_dOH_sfr}
\end{figure*}

The bottom panel of Figure \ref{fig_rad_dOH} presents the median  radial \dOH\ profiles for galaxies at various stages. 
For comparison, the upper panel displays the radial \dssfr\ profiles from \cite{Pan19}. 
The color coding is consistent with that used in Figure \ref{fig_boxplots}. 
A variety of profiles are observed.

During the incoming phase (S1; red curve in the bottom panel), the \dOH\ profile remains relatively stable around zero, indicating a lack of distinct radial \dOH\ dependency. 
This stability is somewhat expected, as galaxies remain distant from their companions at this stage, showing no evident disk perturbation. 
Simulations also predict a very low gas inflow rate during the incoming phase, leading to minimal changes in metallicity \citep{Ion04,Mon10,Per11,Tor12,Bus18}.
Similarly, the impact of galaxy interactions on star formation is not apparent, as the radial sSFR is more or less the same as that of isolated galaxies (red dashed curve in the upper panel).

Simulations predict a substantial increase in gas inflow rate post the initial passage of two galaxies \citep{Ion04,Mon10,Per11,Tor12,Bus18}. 
While direct measurement of gas inflow rate in observations is challenging, metallicity gradient can serve as a proxy.
Our data at this merger stage (S2; yellow curve  in the bottom panel)  displays a noteworthy reduction in \OH\ (negative \dOH) toward the central regions of galaxies. 
The central metallicity is suppressed by  around 0.1 dex compared to the controls.
Beyond the central region, \dOH\ typically increase with radius, but showing widely suppressed compared to the controls across the radial range we explored. 
Specifically, \dOH\ is around $-$0.04 dex at approximately 0.75 \Reff\  and increases to  $\sim$$-$0.01 dex at the outermost radius explored.
Such variation of \dOH\ along the radial direction lead to a flatter \OH\ profile with respect to the controls (Figure \ref{fig_Mstar_OH_slope}).
Furthermore, one can expect a correlation between metallicity and star formation intensity since both are outcomes of gas inflow.
In our previous study, a centrally-peaked starburst is observed during the S2 stage, as seen in the upper panel of Figure \ref{fig_rad_dOH}.
The central sSFR of S2 galaxies experiences a substantial enhancement, reaching  by a factor of four times ($\sim$ 0.6 dex) compared to the controls.

Following the initial passage, the distance between the two interacting galaxies increases. 
The \dOH\ values rise from the S2 to S3 stages (blue curve) across most of the radial range, increasing by $+$0.01 to $+$0.05 dex. 
Simulations incorporating metal enrichment also predict an increase in \OH\ following the first passage, driven by stellar metal enrichment that counteracts the dilution occurring after the starburst event \citep[e.g.,][]{Tor12,Bus18,Tis19,Tis22}.
Comparing the \dOH\ profiles between the S2 and S3 samples shows that the enrichment from the starburst in S2 is more pronounced in the central region compared to the disk region presumably due to the centrally-peaked starburst. 
This results in a reversal of the metallicity gradient observed at the subsequent S3 stage compared to S2. 
Consequently, a negative metallicity gradient, resembling the initial gradient, becomes evident again. 
However, it is important to note that although a negative \OH\ profile is observed for S3, the magnitude of \OH\ is still lower than that of the controls across the entire disk considered.

During the final coalescence phase, when both nuclei are discernible (S4(2); green curve), \dOH\ decreases again, a change attributed to the increase in gas inflow rate \citep[e.g.,][]{Ion04,Tor12,Bus18}. 
Intense, galactic-wide star formation is triggered once more due to the massive gas inflow. 
At this stage, \dOH\ falls below $-$0.05 dex at most radii. Furthermore, the radial dependence of \dOH\ is less pronounced compared to other phases due to the highly disrupted nature of the disk (Figure \ref{fig_merger_stages}). 
Notably, in this context, not only does \dOH\ exhibit such behavior, but the extent of enhanced star formation also lacks distinct radial dependence at this stage.

Finally, the post-merger phase (S4(1); purple curve), \dOH\ consistently decreases presumably due to the peak mass inflow rate at this stage. 
S4(1) galaxies exhibit the lowest \dOH\ levels across all radii explored, with the inner regions showing the most significant metal dilution (\dOH\ $\approx$ $-$0.11 dex), which gradually transitions to less dilution (\dOH\ approximately $-$0.05 dex at 1.5$\mathrm{R_e}$). 
The lowest \dOH\ is associated with the final, and  the most intense, starburst event (Figure \ref{fig_rad_dOH}). 
While simulations predict stellar metal enrichment following the starburst phase, our sample does not exhibit such enrichment at S4(1). 
This discrepancy may stem from the time delay associated with star formation, evolution, and feedback to the ISM. 
Simulations suggest that  metal enrichment occurs approximately 0.2 to 0.5 Gyr after the final merger of the galaxies \citep[e.g.,][]{Tor12,Bus18}. 
During this period, galaxies gradually stabilize, losing visible features such as asymmetry and tidal tails/shells that aid in identifying them as post-mergers by visual inspection \citep[e.g.,][]{Loz08a}. 
Consequently, our S4(1) sample likely represents the phase shortly after the final merger of the two galaxies, and therefore the impacts of stellar feedback and metal enrichment are still progressing.

The fate of a galaxy following a merger (e.g., S4(1)) depends on several factors, including the properties of the progenitor galaxies, their gas fraction, and the dynamics of the interaction. Gas-rich mergers are more likely to reform a disk, as residual gas settles and fuels star formation, leading to the regeneration of a rotationally supported structure \citep{Rob06,Hop09}. In such cases, the metallicity profile may gradually resemble that of an isolated galaxy as star formation and secular-driven radial mixing establish a new equilibrium. In contrast, gas-poor mergers, minor mergers, or those with significant angular momentum loss tend to form early-type galaxies, where violent relaxation and dissipationless stellar mixing dominate the final structure \citep{Bou07,Boi10}. Without sustained star formation to drive chemical enrichment, the metallicity profile may become more homogeneous, with weaker radial gradients compared to disk galaxies \citep{Dim09,San20}. Studying metallicity profiles in merger remnants can reveal whether galaxies return to equilibrium or retain signatures of past interactions. However, identifying merger remnants is challenging, as they gradually lose morphological features over time. Similarly, identifying merger progenitors through observations is also difficult, as interactions can obscure key structural characteristics. In such cases, cosmological simulations play a crucial role in tracking how metallicity profiles evolve and providing insights into the long-term impact of mergers on galaxy evolution \citep[e.g.,][]{Tor19,Tis22,Tap25}.

Figure \ref{fig_OH_vs_SFR} illustrates the relation between \dOH\ and \dssfr\ across different radial bins and merger stages, with darker points indicating regions closer to the galactic center.
Overall, the figure demonstrates an anti-correlation between \dOH\ and \dssfr\ across several stages of the galaxy merger process. This anti-correlation is most pronounced in S2 and S4(1), where the star formation activity is at its peak. In these stages, increased star formation rates are associated with lower oxygen abundances, likely driven by gas inflows and subsequent dilution of metals.

In contrast, for S1 and S4(2), the correlation between \dOH\ and \dssfr\ is less apparent, though for different reasons. In S1, both parameters cluster tightly around zero,  reflecting minimal impact from the galaxy interaction, resulting in negligible variation in star formation and metallicity. Meanwhile, for S4(2), the radial dependence of both \dOH\ and \dssfr\ is obscured by highly perturbed disks, making it difficult to discern clear trends in either parameter.

For S3, there appears to be a positive correlation between \dOH\ and \dssfr, suggesting that, in this stage, the influence of metal enrichment from star formation outweighs the effects of dilution by gas inflows. This could indicate that the galaxy has experienced recent (i.e., at S2), localized star formation, enriching the ISM with metals. 

The results have significant implications for the mass-metallicity-star formation rate (MZR-SFR) relation in galaxies. Our results highlight the dynamic nature of galaxy evolution, in the sense that the MZR-SFR relation is not static but can evolve based on the specific stage of a merger or interaction. These interactions create environments where both gas dilution and metal enrichment processes can occur simultaneously, leading to temporary deviations from the typical mass-metallicity trend. This highlights the importance of accounting for these effects when studying the nature of MZR-SFR.

Finally, although directly measuring gas inflows is challenging, \citet{Gar23}, using data from the APEX/EDGE (molecular gas)-CALIFA (optical IFU) survey, found that \emph{p/m} tend to have higher central molecular gas fractions, serving as indirect evidence of gas inflows. However, in their sample, the inflowing molecular gas does not necessarily have a significant impact on the central oxygen abundance \citep{Gar24}. While this contrasts with other studies, it emphasizes the complexity of galaxy evolution and suggests that factors such as the timing of star formation or feedback processes (from both AGN and star formation) may play key roles in chemical enrichment. Additionally, the sample selection is crucial when comparing results, as the gas inflow $\rightarrow$ abundance dilution $\rightarrow$ star formation $\rightarrow$ feedback $\rightarrow$ abundance enrichment cycle could be a highly dynamic process, potentially varying with the properties of the host galaxy and the nature of the interaction.

\subsubsection{Internal and External Processes Shaping Radial Profiles}
Although this work focuses on the impact of galaxy interactions on metallicity, it is important to note that metallicity gradient flattening can also result from internal processes.
Galactic bars, for example, can redistribute metal-enriched gas through radial flows driven by their non-axisymmetric gravitational potential \citep{Ath92}. This enhances mixing and reduces metallicity variations across the disk. Observations and simulations suggest that barred galaxies often exhibit shallower metallicity gradients than unbarred ones \citep[e.g.,][]{Mar97,Dut99}, though not always \citep[e.g.,][]{Zur21,Gra22,Che23}. This supports the idea that secular processes, such as bar-driven inflows and turbulence, can flatten metallicity gradients independently of mergers. Our sample includes barred galaxies, making it possible that bars also contribute to the observed metallicity trends.

At the same time, galaxy mergers can also contribute to bar formation \citep{Moe17, Cav20}. Studies show that mergers with low mass ratios and specific alignments can destabilize the disk, triggering bar formation through gravitational interactions. However, distinguishing the effects of mergers from those of secular processes, such as internal disk instabilities and radial gas flows, is challenging due to the complex and intertwined nature of these mechanisms. A large statistical sample, such as the full MaNGA dataset, could provide further insights into these interactions and help disentangle their relative contributions to metallicity gradient evolution.

In addition, feedback processes, such as star formation-driven large-scale winds and supernova explosions, can impact metallicity distribution. These energetic outflows expel metal-enriched gas from the central regions and redistribute it across the galaxy, altering metallicity gradients. Depending on their efficiency and direction, these winds can either flatten or steepen the gradient. While such processes are ubiquitous at high redshifts \citep{New12,For19}, they are also relevant in galaxies with high star formation rates or recent bursts of star formation, as seen in our sample.

\subsection{\dOH\ versus Projected Separation}
\label{sec_oh_rp}
Previous observational studies often use the projected separation \rp\ between two galaxies as an indicator of their merger stage. 
Building on this approach, Figure \ref{fig_rp} analyzes the distribution of \dOH\ in spaxels across different regions of galaxies, focusing specifically on variations between the central regions ($R$ $<$ 0.5~\Reff; panel (a)) and the disk regions (0.5 $<$ $R$ $<$ 1.5~\Reff; panel (b)). 
The spaxels are further categorized into four distance ranges based on the \rp\ of the host galaxy: greater than 40 kpc, between 20 and 40 kpc, and less than 20 kpc, along with the final coalescence and post-merger phases where the \rp\ are not determinable. 
For comparative purposes, the \dOH\ distribution of the control galaxies is also presented.
The yellow circles  mark the median \dOH\ values in each distribution.

The central regions typically exhibit more significant fluctuations in metallicity distribution compared to the disk regions. 
An excess of lower \dOH\ at around $-$0.2 dex is observed in the central distributions across all \rp\ groups. 
This excess is less pronounced in the disk regions, implying a higher sensitivity to external influences like gas inflow or outflow, or interaction-induced starbursts in the galaxy cores. 
In contrast, the disk regions might reflect a more diluted or delayed response to similar influences, possibly due to a broader distribution of effects and a more gradual mixing of ISM materials.

Moreover, Figure \ref{fig_rp} shows that closer proximities to other galaxies (\rp\ $<$ 20 kpc) appear to have more pronounced changes in metallicity, indicative of interaction-driven processes.
This effect is observed in both central and disk regions, but it is particularly significant in the central areas. 
In these central regions, the median \dOH\ shows a marked decreasing trend, moving from nearly 0 dex in the largest separation bins to $\sim$ $-$0.08 dex during the final coalescence and post-merger phases. 
This is consistent with the findings from \cite{Scu12}, where nuclear \dOH\ is measured using single-fiber spectroscopy from SDSS, as well as from simulations \citep{Bus18}.
Meanwhile, the disk regions exhibit a subtler decrease in \dOH, declining from close to 0 dex to $\sim$ $-$0.06 dex during the same phases.

However, despite their relevance, projected separation is not a direct or linear indicator of merger stages. 
This is because two galaxies often undergo several pericenter passages before finally merging, as demonstrated in simulations \citep[e.g.,][]{Tor12, Bus18}. 
For instance, the \dOH\ distributions of galaxy pairs that are widely separated (\rp\ $>$ 40~kpc) actually represent a mix of galaxies both before (S1) and after (S3) the first pericenter passage. 
In fact, the observed excess of lower \dOH\ at \rp\ $>$ 40~kpc in both central and disk regions is primarily attributed to galaxies after the first pericenter passage (S3).
If we were to consider only the galaxies at S1, their \dOH\ distributions would remarkably resemble those of the control groups (as seen in Figure \ref{fig_boxplots}), indicating minimal deviation from unperturbed states. 
Such a mix of multiple stages of mergers occurs in other \rp\ bins as well.
This highlights the complexity of using projected separation as a metric for understanding galactic interactions and highlight the necessity of considering the dynamical history of galaxy pairs to accurately interpret their merger stages and associated chemical (this work), as well as star formation \citep{Pan19} signatures.

On the other hand, using the visualization method in Figure \ref{fig_rp}, we observe that many spaxels, regardless of their \rp\ or location within the galaxies, exhibit \emph{unaffected} \OH, with \dOH\ remaining around zero. This population, which appears resistant to changes in metallicity, is significant and will be discussed in detail in the next section.

\subsection{Spaxels with Normal \OH}
\label{sec_oh_normal}
\begin{figure*}
	\centering
	\subfigure[]{\label{fig_rp_center}\includegraphics[width=0.75\textwidth]{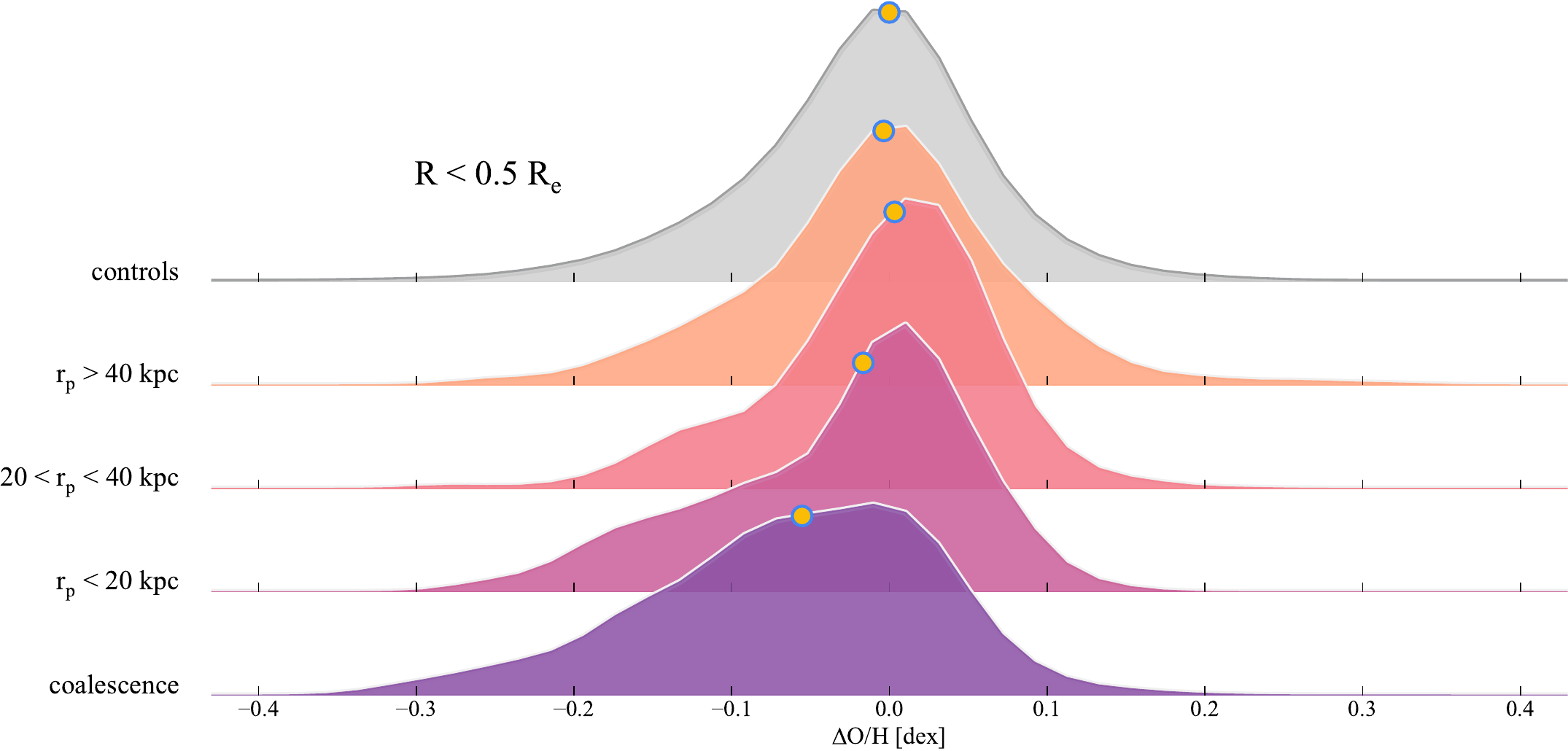}}
	\subfigure[]{\label{fig_rp_disk}\includegraphics[width=0.75\textwidth]{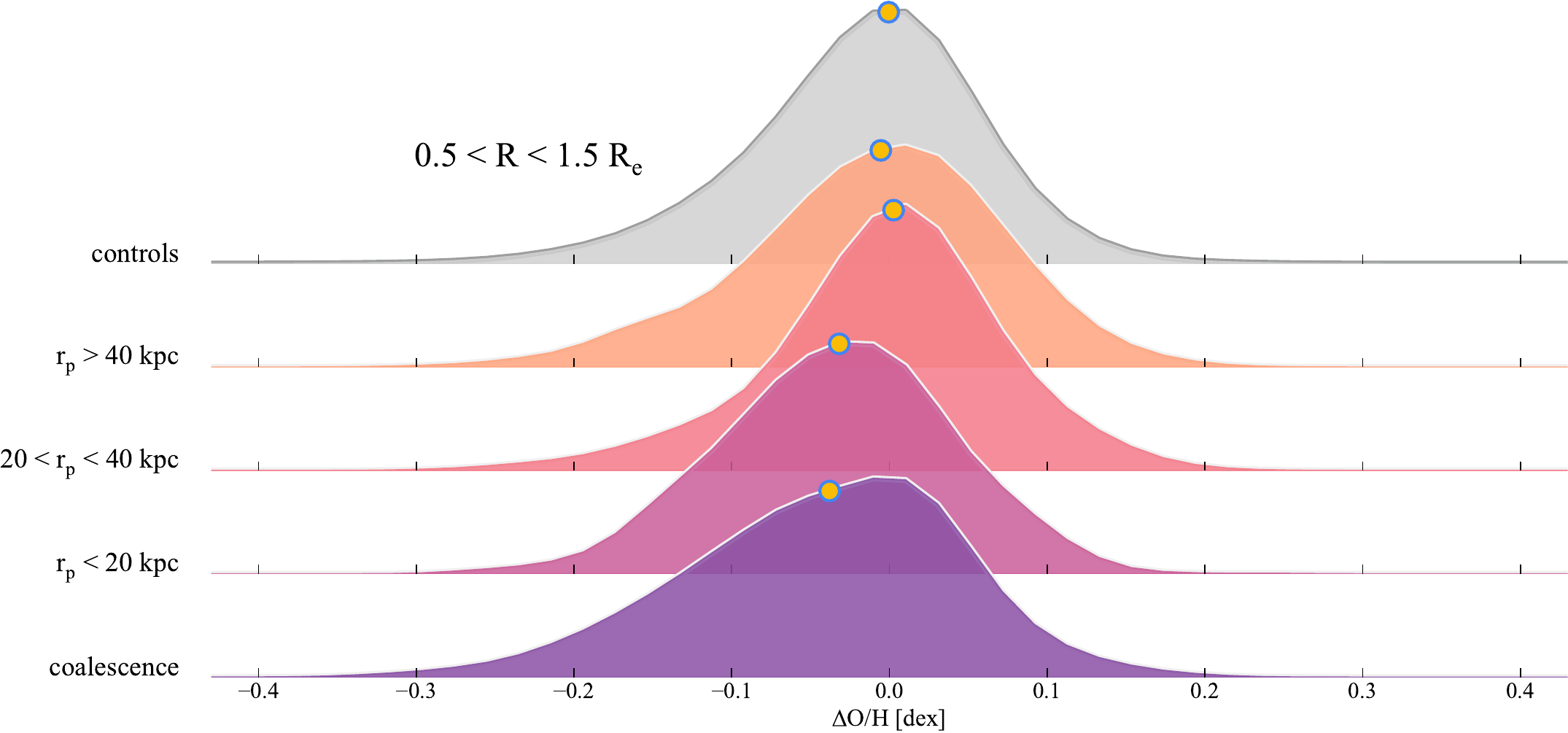}}
	\caption{Distribution of \dOH\ for different galaxy groups separated by projected pair separation (\rp) and post-mergers. Panel (a) shows the distributions of spaxels within the inner region ($R$ $<$ 0.5 \Reff), while panel (b) shows the distributions for the outer spaxels (0.5 $<$ $R$ $<$ 1.5 \Reff). The groups include control galaxies (gray), galaxies with \rp\ $>$ 40 kpc (orange), 20 $<$ \rp\ $<$ 40 kpc (red), \rp\ $<$ 20 kpc (pink), and in final coalescence phase (purple). The yellow dots indicate the median \dOH\  for each group. Metallicity progressively decreases as galaxies approach each other; however, the central regions generally exhibit a greater degree of metallicity dilution compared to the disk regions based on median values.
	}  
	\label{fig_rp}
\end{figure*}

The \dOH\ distributions in Figures  \ref{fig_boxplots} and \ref{fig_rp} indicate that while galaxy interactions and mergers can reduce  \dOH\ in galaxies, the effect is not uniform across all spaxels. 
Many spaxels still exhibit  \dOH\ values close to those of their controls in the data.
Tracking the variation of \dOH\ over time remains challenging based on observations. 
Consequently, it is not evident whether the \dOH\ values of these spaxels remain unchanged during the interaction process or have undergone a balance of gas dilution and enrichment.

Moreover, the impact of interactions on metallicity might not be uniformly distributed across a galaxy. 
Certain regions might experience more significant changes due to direct interaction effects such as tidal forces or shocks, while others might remain largely unaffected  \citep[e.g.,][]{Hwa19}. 

In addition, the chemical signatures of interaction-induced changes   might not be immediately observable. 
There could be a time lag between the onset of interaction effects (e.g., gas inflow and metal dilution), the triggering of star formation (metal enrichment) and the detectable changes in \OH\ \citep{Tor12,Per11} due to slow diffusion processes, the lifespan of massive stars, and the delay in the release of metals from supernovae \citep{Mon10}.

Finally, the configuration of interacting galaxies can significantly influence their metallicity evolution \citep{Lin08,Per11,Ath16}. 
In wet mergers, where both galaxies are gas-rich and have clumpier ISM, intense starbursts can be triggered by the compression of gas. 
Consequently, metallicity variations, including both dilution and enrichment, are presumably more  pronounced. 
In contrast, gas-poor (dry) mergers lack readily available gas, making it challenging to initiate starbursts and, consequently, to observe significant metallicity variations.
Our sample size is currently too small to further classify the merger configuration across different merger stages. 
Future larger datasets, along with more efficient merger and morphology classification methods  -- such as those employing machine learning techniques \citep[e.g.,][]{Cha22,Dom22}  -- will be crucial for exploring and quantifying the impact of merger configurations on star formation and metallicity evolution during galaxy interactions.

Overall, the distributions of \dOH\ in spaxels within galaxies in pairs and mergers  highlight the complex interplay of physical processes that govern galactic evolution during interactions. 
The specific outcome in terms of metallicity changes can vary widely depending on the initial conditions, the nature of the interaction, and the timescale over which observations are made.
The observed distributions of  metallicity variations of spaxels in pairs and mergers in this study  provide a critical empirical baseline that can significantly refine the accuracy of simulation models of metal enrichment, star formation, and galaxy interactions and evolution.

\subsection{Caveats}
A few important considerations should be kept in mind when interpreting the results.
First of all, it is important to note that our discussion on radial profiles is based on the \emph{median profiles} of galaxies at different merger stages. 
However, individual galaxies can exhibit a wide range of radial profiles (see \citet{Tho19} as an example for \emph{p/m} and \citet{Per16} for galaxies in general). 
This diversity likely arises from variations in galaxy morphologies and merger configurations, influenced by factors such as mass ratio, progenitor properties (e.g., gas fraction), and merger orbits. Simulations have tested some of these parameters and confirmed their impact on the distribution of star formation and metallicity \citep[e.g.,][]{Tor12,Mor21}.
These variations emphasize the complexity of galaxy interactions and the necessity of considering individual characteristics when interpreting radial metallicity profiles.

Moreover, this work focuses on oxygen abundance. However, it is well known that the production and diffusion timescales of oxygen are much shorter than those of other metals, such as iron-peak elements (e.g., Fe and Ni) \citep[see the review by][]{Mao19}. A natural direction for future work would be to compare the radial profiles of different elements, which would offer a powerful tool for constraining the evolution of star formation histories in galaxies undergoing interactions and mergers.

Additionally, the anti-correlation between \dOH\ and \dssfr\ observed in certain merger stages is consistent with established relations, such as the MZR-SFR relation, and points to a meaningful physical connection between gas inflows, star formation, and metal production and enrichment. However, the differences in absolute O/H scales are subtle and fall within the typical uncertainties of the calibrators used. Moreover, secondary factors, such as the ionization parameter (U) and variables like the age of ionizing stars, could influence both O/H and sSFR measurements. We are aware of these complexities and will address them in greater detail in future work.

Deriving the \Reff\ for galaxies in pairs and mergers can be inherently challenging due to the complex morphologies and disturbances caused by interactions. This complexity can introduce potential uncertainties in our radial profiles. Similarly, while the radial profiles in this work have been de-projected based on the disk inclination, determining galaxy inclination -- particularly for the disturbed pair and merger population -- can also be difficult. This introduces another layer of uncertainty to our radial profiles.

Finally, our merger stage classification may introduce additional uncertainties. This issue was discussed and tested in our previous study using the same dataset \citep{Pan19}, and we summarize the key points here.
A major concern is that galaxies lacking clear morphological distortions or asymmetries may be misclassified due to sensitivity limits, as deeper imaging could reveal hidden interaction features. However, statistical tests on redshift distributions suggest no significant bias related to sensitivity \citep[see][]{Pan19}. Additionally, distinguishing flybys from direct mergers remains challenging, as flybys can trigger star formation and gas inflow without leading to coalescence. Flybys are relatively common, contaminating galaxy-pair samples at a 20 -- 30\% level \citep{Sin12}, making their effects difficult to assess. Addressing these complexities requires deeper observations, large statistical samples, and dedicated simulations.

\section{Summary}
\label{sec_summary}
This study explores the spatial evolution of gas-phase metallicity in galaxies in pairs and mergers using data from the SDSS-IV MaNGA survey. 
The IFU observations enable us to obtain spatially-resolved metallicity measurements up to a radius of 1.5~\Reff\ (Section \ref{sec_data}).
By comparing galaxies in pairs and mergers with isolated controls (Figure \ref{fig_controls_oh}), we quantify characteristic trends in how galaxy interactions modify metallicity gradients over time, ranging from early encounters to final coalescence (Figure \ref{fig_merger_stages}).  
Our primary findings can be summarized as follows:

\begin{itemize}
	\item  \emph{Metallicity values.} The early approach of two galaxies has minimal impact on metallicity. However, at and after the first pericenter passage, galaxies in pairs and mergers statistically exhibit lower gas-phase metallicity compared to their isolated controls (Section \ref{sec_spaxel_distribution} and Figure \ref{fig_main}).
	
	\item \emph{Metallicity profiles.} We observe that metallicity gradients generally tend to flatten shortly after the first pericenter passage, likely due to radial gas mixing. However, this flattening is not consistent across all cases and can vary depending on the intensity of the interaction and the level of star formation activity. As the interaction progresses through subsequent stages, the degree of metallicity enrichment or dilution varies, leading to gradients that can either flatten or steepen compared to the initial profiles. Our findings suggest that the common assumptions about the uniform impact of interactions on metallicity profiles may need to be reconsidered (Section \ref{sec_spatial_distribution} and Figure \ref{fig_rad_dOH}).
	
	\item \emph{Metallicity profiles vs. star formation profiles.} The  metallicity and star formation profiles in galaxies undergoing interactions reveals a notable anti-correlation. Regions with lower metallicity are often associated with enhanced star formation, while areas with higher metallicity tend to show reduced star formation activity. This pattern, along with the evolution of metallicity and star formation across different merger stages, suggests a strong connection between gas inflow, which triggers star formation, and the subsequent metal enrichment driven by stellar feedback (Section \ref{sec_spatial_distribution} and Figure \ref{fig_rad_dOH}).

	\item \emph{Metallicity vs. projected separation.} We also study the spaxel-based metallicity changes as a function of the \rp\ between galaxy pairs. The analysis indicates that closer galaxy proximities correlate with more significant changes in metallicity, particularly in central regions. The distribution of metallicity within a given \rp\  range is complex, reflecting the challenges of using \rp\  as an indicator of merger stages due to the varied interaction histories and multiple pericenter passages that galaxies undergo before the final coalescence. This complexity highlights the need to consider the dynamical history of galaxy pairs for accurate interpretation of their merger stages and associated chemical signatures (Section \ref{sec_oh_rp}, \ref{sec_oh_normal}, and Figure \ref{fig_rp}).
\end{itemize}

Our findings shed new light on the complex processes reshaping galaxies during interactions, highlighting the interplay between gas flows, star formation, and feedback mechanisms. 
The detailed spatial patterns of metallicity evolution observed provide crucial empirical data to inform and refine future simulation models of galaxy interactions and evolution.

In addition, while this study does not directly measure gas inflows, which would provide a more definitive link between star formation and metallicity dilution or enrichment, future work will also focus on a detailed kinematic analysis to model gas flows in merging systems (C. L\'opez-Cob\'a in preperation). By examining different phases of the ISM, including ionized and cold neutral gas, we aim to better understand how gas dynamics drive chemical evolution during interactions.


We sincerely thank the anonymous referee for their constructive feedback, which has helped improve the clarity of this manuscript.
This work is supported by the National Science and Technology Council of Taiwan under grants 110-2112-M-032-020-MY3 and 113-2112-M-032 -014 -MY3.
LL acknowledges the National Science and Technology Council of Taiwan for providing support through the grant NSTC 113-2112-M-001-006-.

Funding for the Sloan Digital Sky Survey IV has been provided by the Alfred P. Sloan Foundation, the U.S. Department of Energy Office of Science, and the Participating Institutions. SDSS-IV acknowledges support and resources from the Center for High-Performance Computing at the University of Utah. The SDSS website is www.sdss.org.
SDSS-IV is managed by the Astrophysical Research Consortium for the Participating Institutions of the SDSS Collaboration including the Brazilian Participation Group, the Carnegie Institution for Science, Carnegie Mellon University, the Chilean Participation Group, the French Participation Group, Harvard-Smithsonian Center for Astrophysics, Instituto de Astrofísica de Canarias, The Johns Hopkins University, Kavli Institute for the Physics and Mathematics of the Universe (IPMU)/University of Tokyo, Lawrence Berkeley National Laboratory, Leibniz Institut f{\"u}r Astrophysik Potsdam (AIP), Max-Planck-Institut f{\"u}r Astronomie (MPIA Heidelberg), Max-Planck-Institut f{\"u}r Astrophysik (MPA Garching), Max-Planck-Institut f{\"u}r Extraterrestrische Physik (MPE), National Astronomical Observatory of China, New Mexico State University, New York University, University of Notre Dame, Observatário Nacional/MCTI, The Ohio State University, Pennsylvania State University, Shanghai Astronomical Observatory, United Kingdom Participation Group, Universidad Nacional Autónoma de México, University of Arizona, University of Colorado Boulder, University of Oxford, University of Portsmouth, University of Utah, University of Virginia, University of Washington, University of Wisconsin, Vanderbilt University, and Yale University.

\appendix

\section{Metallicity Calibrators}
\label{appendix_oh_cal}
\setcounter{figure}{0}
\renewcommand{\thefigure}{A\arabic{figure}}

We evaluate the robustness of our main findings presented in Figure \ref{fig_rad_dOH} by comparing them with results obtained using different metallicity calibrations. 
Our default calibration is \emph{O3N2} by \cite{Mar13}. 
Additionally, we explore the commonly used \emph{N2S2} calibration from \citet{Dop16} and the \emph{N2} calibration  from \citet{Mar13}. 
The \emph{N2S2} ([\ion{N}{2}]/[\ion{S}{2}]) and \emph{N2} ([\ion{N}{2}]/H$\alpha$) calibrations  employ photoionization models to convert observed line ratios into metallicity estimates\footnote{Note that all the adopted calibrations are $N_2$-based. The non-$N_2$-based calibration, $R_{23}$ ($\frac{[\text{O II}] \lambda 3727 + [\text{O III}] \lambda\lambda 4959, 5007}{\text{H}\beta}$), is also a common metallicity tracer, but its use comes with two major caveats. First, $R_{23}$ has a double-valued nature, meaning it follows two branches (low-metallicity and high-metallicity), requiring an additional diagnostic to determine the correct metallicity regime, which can introduce uncertainties. Second, $R_{23}$ is sensitive to the ionization parameter, which depends on the excitation conditions of \ion{H}{2} regions \citep[see][]{Kew02}. This makes $R_{23}$ less reliable in environments with varying ionization conditions, such as interacting galaxies. Given these limitations, we have opted to use $N_2$-based calibrations throughout this work.}.
These methods are advantageous because they only require two emission lines that are closely spaced in wavelength, making them less affected by wavelength-dependent dust reddening. 
Additionally, these calibrations are robust against variations in ionization parameter and pressure, making them also widely used in the analysis of gas-phase metallicity in galaxies \citep[see the review by][]{Kew19}.

\begin{figure}
	\centering
	\includegraphics[scale=0.6]{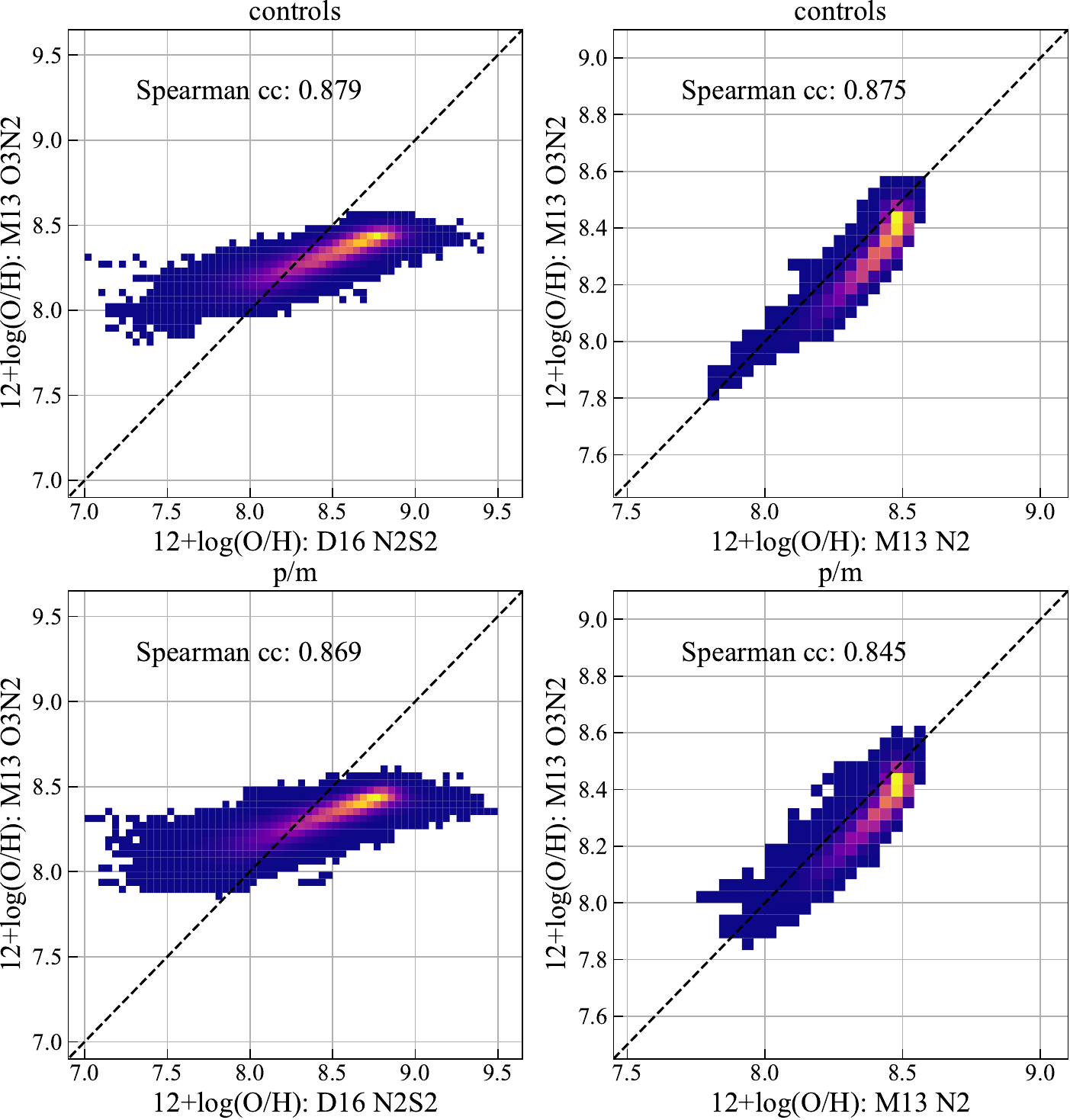}
	\caption{Spaxel-by-spaxel comparisons of different metallicity calibrations. The top row displays comparisons between \emph{O3N2}-based and \emph{N2S2}-based calibrations (left), and \emph{O3N2}-based and \emph{N2}-based calibrations (right) for the control sample. The bottom row shows the comparisons for galaxies in pairs and mergers, arranged in the same sequence. Spearman correlation coefficients between each pair of metallicity calibrators are displayed in the upper-left corner of each panel. The dashed line represents the one-to-one correlation, providing a reference for direct comparison between the different calibrations. The color scale indicates the density of data points, with lighter colors representing higher density. Metallicity derived from different calibrations are in qualitative agreement with each other.
	}  
	\label{fig_metallicity_comparison}
\end{figure}

Figure \ref{fig_metallicity_comparison} presents spaxel-based comparisons of metallicity using the \emph{O3N2} method against those obtained with \emph{N2S2} (left column) and \emph{N2} (right column). The top two panels display comparisons for the control sample, whereas the bottom two panels focus on galaxies in pairs and mergers.
The Spearman correlation coefficient between each pair of metallicity calibrators is shown in the upper-left corner of each panel.
The metallicity determined using the \emph{O3N2} method shows a positive correlation with those derived from \emph{N2S2} and \emph{N2}, indicating that the three calibrations are qualitatively consistent.
This correlation is nearly consistent for both the control sample and the sample of galaxies in pairs and mergers.

While metallicity values derived from the \emph{O3N2} and \emph{N2} calibrators generally agree, those based on the \emph{N2S2} method cover a broader range compared to  \emph{O3N2}, yet still show a strong positive correlation as indicated by the correlation coefficient.
Discrepancies similar to those observed here have also been reported in previous studies \citep[e.g.,][]{Scu21}. 
Exploring the root causes of these discrepancies, however, exceeds the goal of this paper. 
Our focus instead lies on examining whether systematic variations among different metallicity calibrations could influence the results of this study.
For further insights into comparisons and conversions between various metallicity calibrations, we refer the reader to the works of \citet{Kew08}, \citet{Scu21}, and \citet{Tei21} as examples.

\begin{figure*}
	\begin{center}
		\subfigure[]{\label{fig_rad_OH_N2S2}\includegraphics[width=0.48\textwidth]{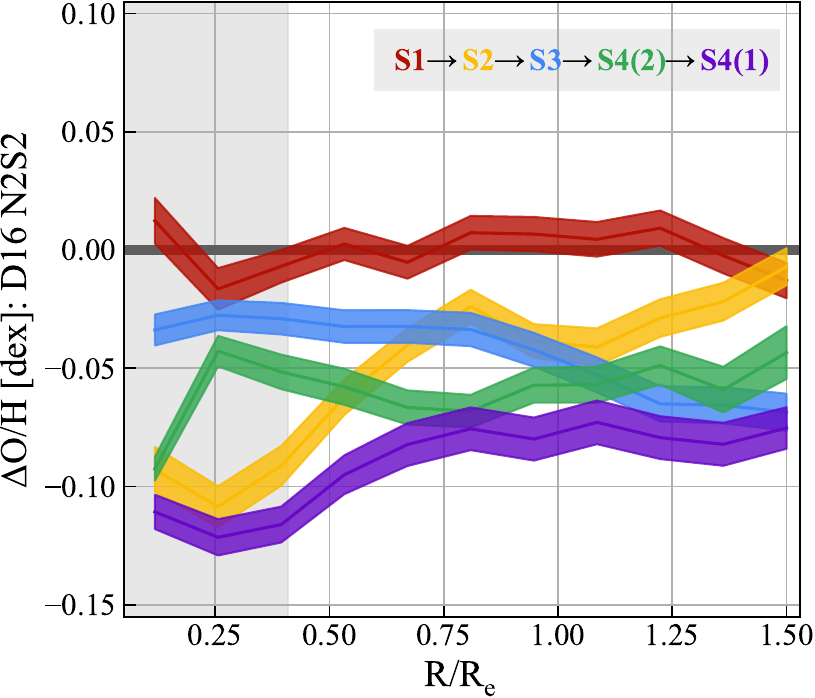}}
		\subfigure[]{\label{fig_rad_OH_N2}\includegraphics[width=0.48\textwidth]{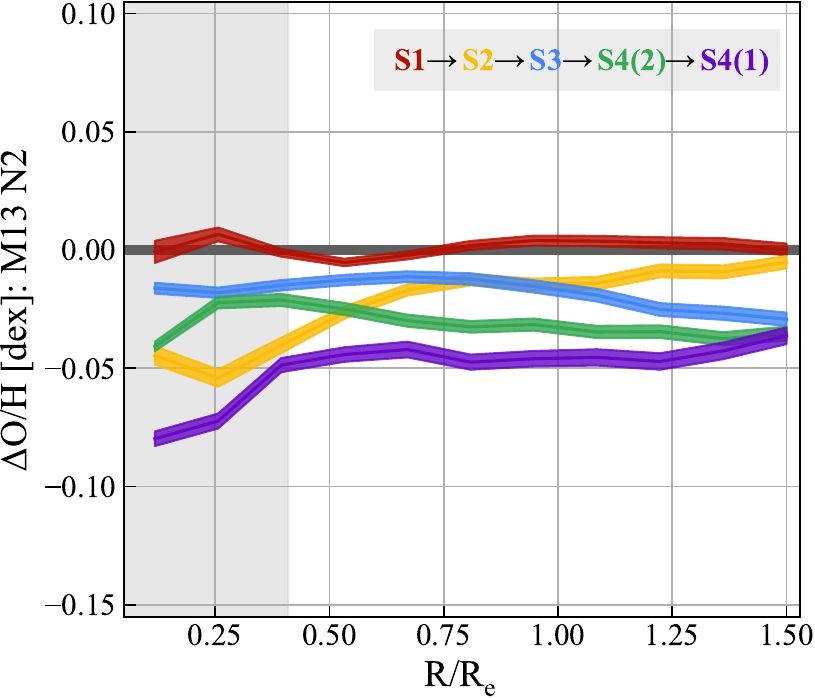}}
	\end{center}
	\caption{Radial profiles of \dOH\ for galaxies at various merger stages. The left and right panels show the radial \dOH\ profiles based on the \emph{N2S2} and \emph{N2} calibrations, respectively. The color coding for the different merger stages corresponds to that used in Figure \ref{fig_rad_dOH}. The characteristics of radial \dOH\ across different merger stages, as identified in Figure \ref{fig_rad_dOH} using the default calibration, are also evident in these alternative calibrations.
	}
	\label{fig_rad_OH_other_calibrations}
\end{figure*}

Figures \ref{fig_rad_OH_other_calibrations}(a) and \ref{fig_rad_OH_other_calibrations}(b) show the radial profiles of \emph{N2S2}-based and \emph{N2}-based \dOH\ across galaxies at various merger stages. The \emph{N2S2}-based \dOH\ generally spans a broader range compared to both the \emph{N2}-based and the default \emph{O3N2}-based \dOH\ (see Figure \ref{fig_rad_dOH}) across all radii. This broader range is a natural result of the systematic differences between the various metallicity calibrators, as illustrated in Figure \ref{fig_metallicity_comparison}.

Overall, we observe that the characteristics of radial \dOH\ across different merger stages, as identified in the default calibration, are also reflected in other calibrations.
Here, we highlight the key characteristics that are reproduced in the \emph{N2S2}-based and \emph{N2}-based metallicity analyses.
The profiles of \emph{S1} fluctuate around the zero point across all radii. 
Meanwhile, the profiles of \emph{S2} exhibit a significant decline towards the central regions, reaching the second lowest value after that of stage \emph{S4(2)}.
Subsequently, \dOH\ increases from \emph{S2} to \emph{S3}, with the \emph{S3} galaxies demonstrating statistically higher \dOH\ in the inner regions compared to the outer regions, leading to a slope that is inverse to that observed in \emph{S2}.
The radial profiles of \emph{S4(1)} exhibit no discernible radial dependence, likely attributable to the chaotic nature of disks at this stage.
Finally, the magnitude of \dOH\ decreases to its lowest value across nearly all radii in \emph{S4(2)}, with the most significant drops observed in the central regions.

Based on the analysis presented in this section, we conclude that our primary findings demonstrate robustness against the selection of commonly used metallicity calibrations. 
Despite the inherent differences between various calibration methods \citep[e.g.,][]{Kew08,Scu21,Tei21}, the  conclusions drawn from this work remain consistent and reliable.

\section{Matching Scheme}
\label{appendix_matching}
\setcounter{figure}{0}
\renewcommand{\thefigure}{B\arabic{figure}}

The spaxels, no matter in the isolated or \ppm\ samples, are matched with a set of control spaxels selected based on the global properties of the host galaxies ($z$, \Mstar, and size) and the local properties ($\Sigma_{\ast}$ and position).
Since we are focused on spatially-resolved \OH\ properties, matching spaxels based on local characteristics is a logical approach. However, the importance of also matching in global galaxy properties may not be as immediately apparent. 
The rationale for matching spaxels based on global galaxy properties is discussed in this section.
 
\begin{figure*}
	\begin{center}
		\subfigure[]{\label{fig_scatter_resolved_Z_Mstar}\includegraphics[width=0.48\textwidth]{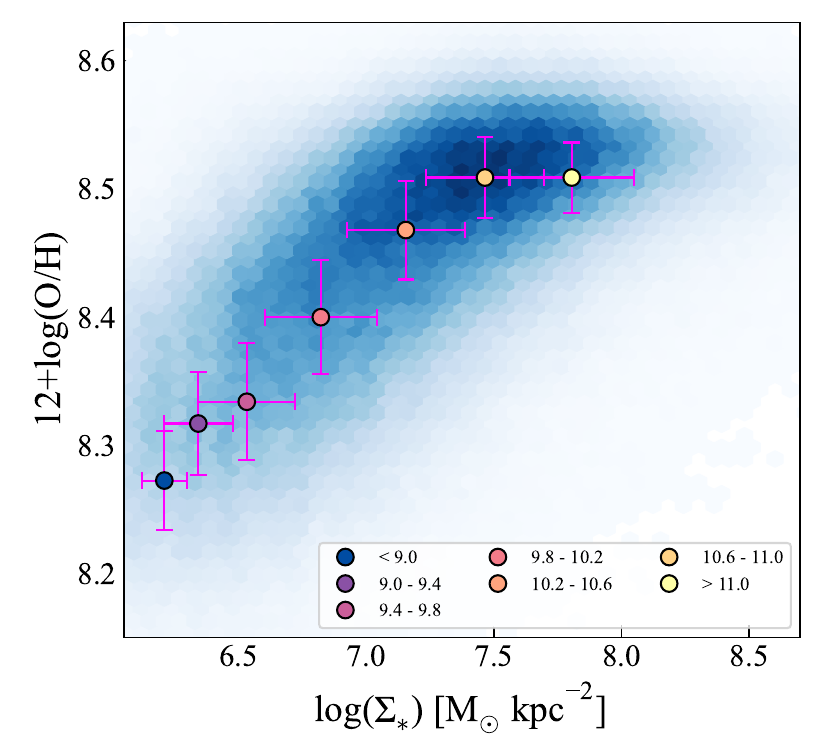}}
		\subfigure[]{\label{fig_scatter_resolved_Z_Mstar2}\includegraphics[width=0.48\textwidth]{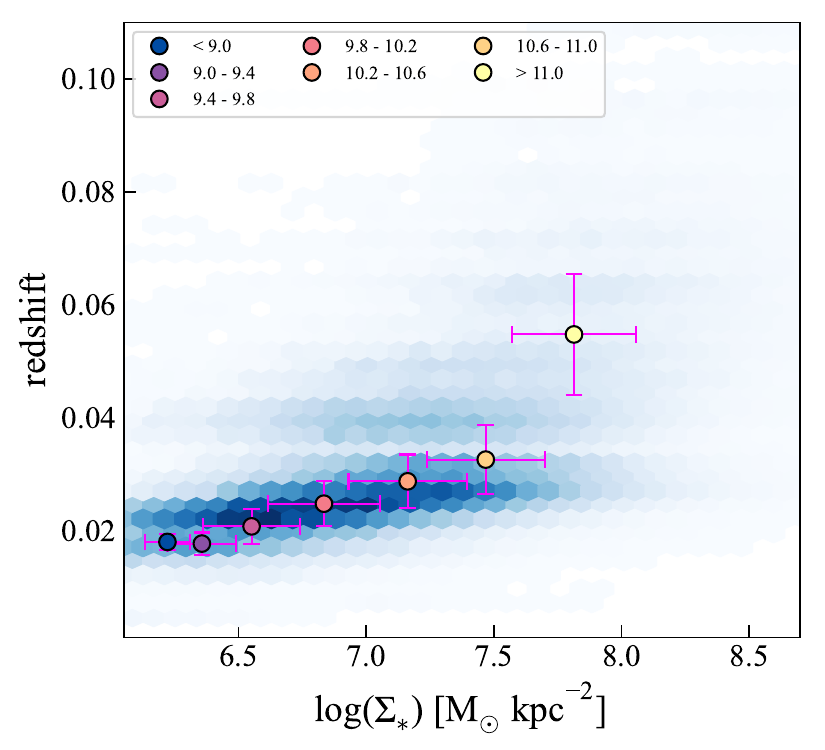}}
	\end{center}
	\caption{(a) Metallicity as a function of $\Sigma_{\ast}$. The hexagonal bins represent the density of spaxels, with darker blue regions indicating higher spaxel densities. The colored points with error bars show the median metallicity values for different global log($M_{\ast}/M_{\sun}$) bins. The figure demonstrates a clear positive correlation between metallicity and $\Sigma_{\ast}$. This trend is consistent across different global \Mstar\ bins. The figure is constructed using spaxels from the control sample. 
		(b) Redshift as a function of $\Sigma_{\ast}$. The colors and symbols are the same as in panel (a). There is a noticeable correlation between $\Sigma_{\ast}$ (and consequently \Mstar) and redshift, a result of the MaNGA sample design. This dependency ensures consistent spatial resolution across galaxies with different masses.
	}
	\label{fig_scatter_sm_z_oh}
\end{figure*}

In the local Universe, galaxies exhibit a positive correlation between their global \Mstar\ and metal content, a relationship known as the Mass--Metallicity Relation (MZR; \citealt{Tre04}). 
This relation indicates that, on average, more massive galaxies are more metal-rich than their less massive counterparts. 
Originally, the MZR was established using integrated measurements of entire galaxies. 
However, recent advancements have extended this relation to spatially-resolved studies (\citealt{Ros12}; \citealt{San14}, and see Figure \ref{fig_scatter_sm_z_oh}(a) in \citealt{Bar16}  as a demonstration by  using MaNGA data), allowing for a more detailed understanding of how metallicity varies within galaxies as a function of their mass and the origin of this relationship.
Therefore, to isolate the effects of galaxy interactions, it is necessary to account for this intrinsic correlation. 
This is achieved by matching the spaxel in question with a similar $\Sigma_{\ast}$.
Furthermore, $\Sigma_{\ast}$ is radially dependent, generally decreasing outward. 
While the radial profiles of $\Sigma_{\ast}$ tend to be uniform across galaxies, regardless of morphology and mass \citep[e.g.,][]{San20}, the magnitude of $\Sigma_{\ast}$ varies as a natural consequence of its relationship with global \Mstar\ (Figure \ref{fig_scatter_sm_z_oh}(a)). 
Therefore, to ensure accurate matching, it is essential to consider global \Mstar\ as well in the matching scheme.

For each bin of global \Mstar, the MaNGA target selection defines a minimum and maximum redshift range. 
This strategy ensures that the sample maintains a consistent angular size distribution across all \Mstar\ bins. 
Consequently, this allows us to probe spatial resolution in units of effective radius uniformly across all mass ranges. 
However, as a result of this approach, galaxies with higher \Mstar\ tend to be located higher redshifts, and vice versa. 
This unavoidable bias is a byproduct of the need to maintain consistent spatial resolution across the sample.
Further details on the MaNGA sample selection are presented in \cite{Wak17}.

This \Mstar--redshift relationship also holds at the local level, leading to a correlation between $\Sigma_{\ast}$ and redshift in our reference sample, as shown in Figure \ref{fig_scatter_sm_z_oh}(b). 
When matching a spaxel with controls based on \Mstar\ and $\Sigma_{\ast}$, redshift is indirectly considered due to this mass--redshift dependency. 
However, Figure \ref{fig_scatter_sm_z_oh}(b) indicates that the redshift scatter within each $\Sigma_{\ast}$ or \Mstar\ bin is not negligible. 
To avoid any redshift-related biases in metallicity properties, we explicitly include redshift as a parameter in our matching scheme. 

\begin{figure*}
	\begin{center}
		\subfigure[]{\label{fig_S1}\includegraphics[width=0.32\textwidth]{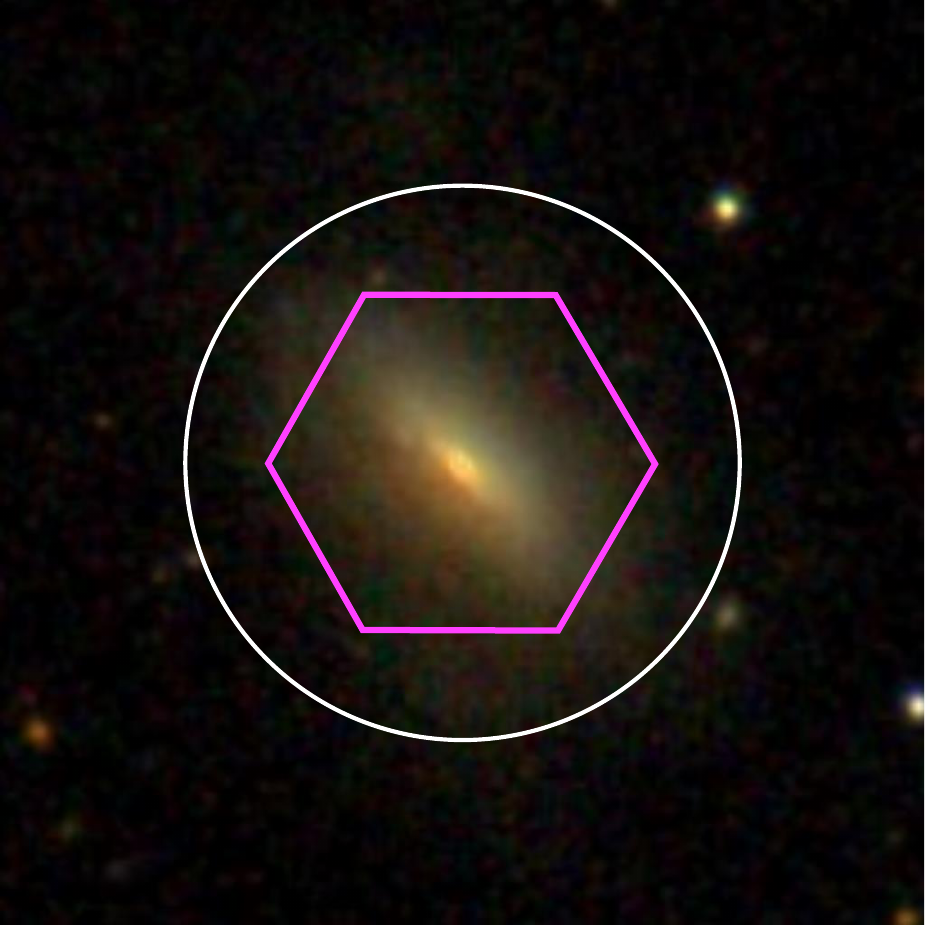}}
		\subfigure[]{\label{fig_S2}\includegraphics[width=0.32\textwidth]{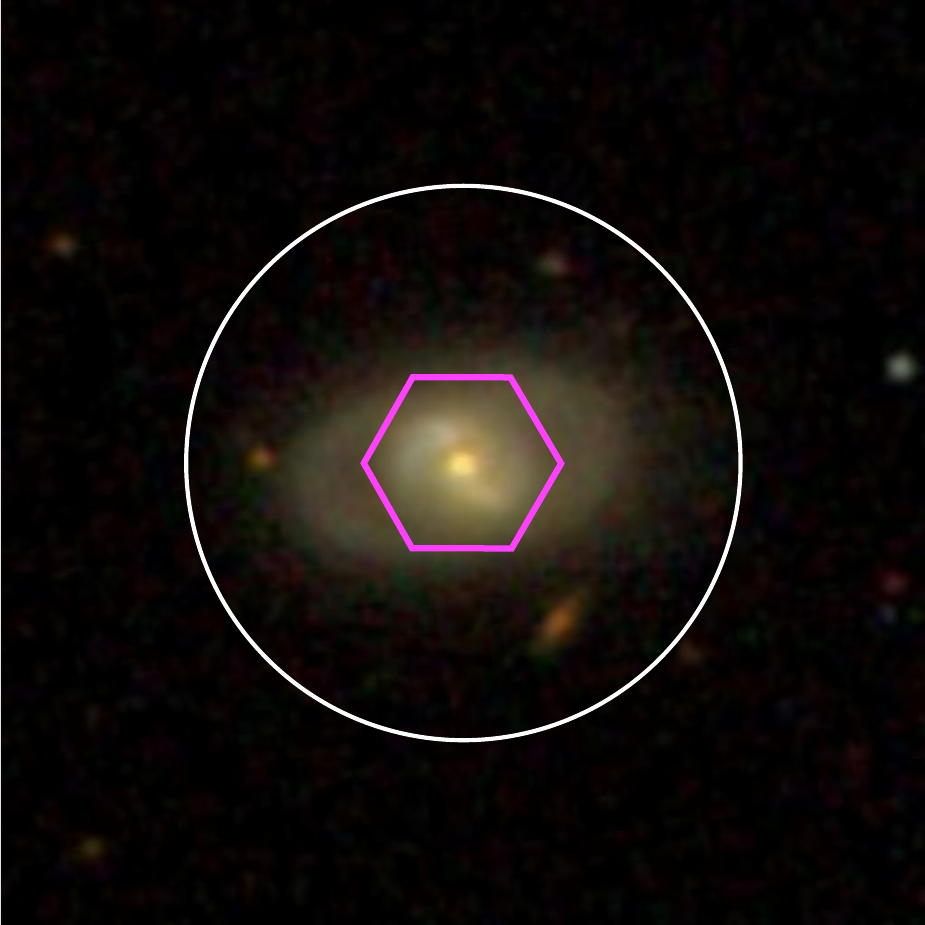}}
		\subfigure[]{\label{fig_S3}\includegraphics[width=0.32\textwidth]{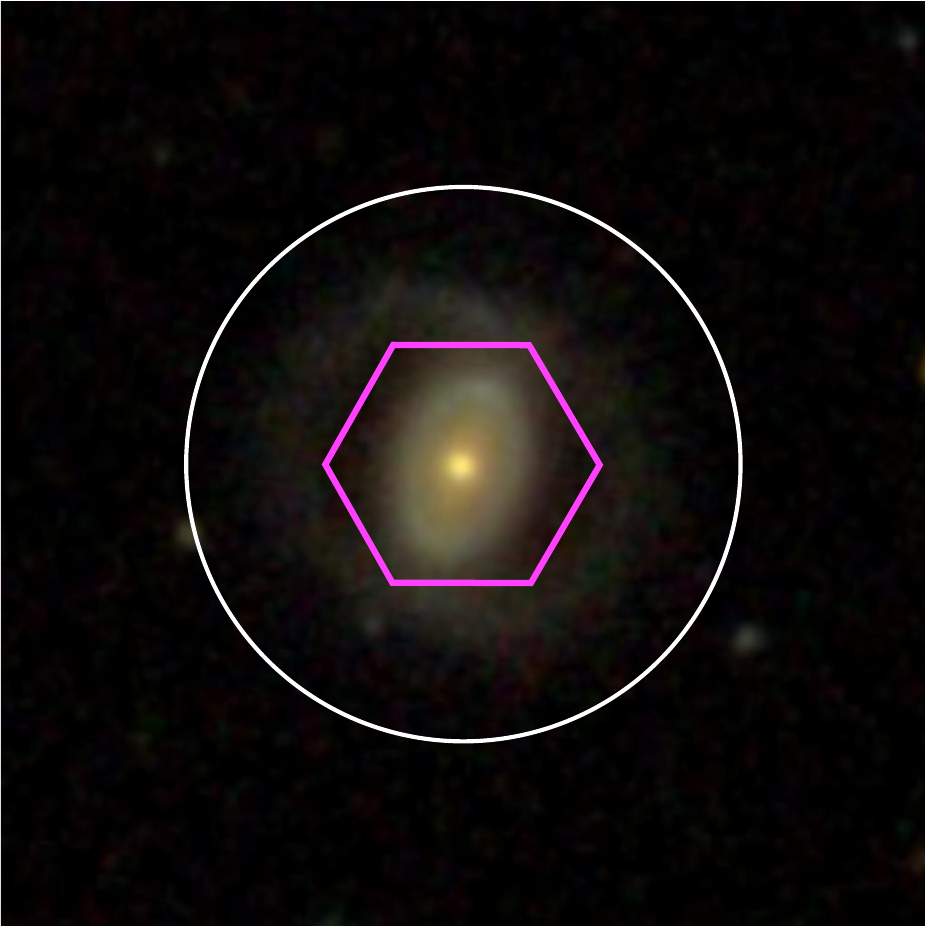}}
		
		\subfigure[]{\label{fig_L1}\includegraphics[width=0.32\textwidth]{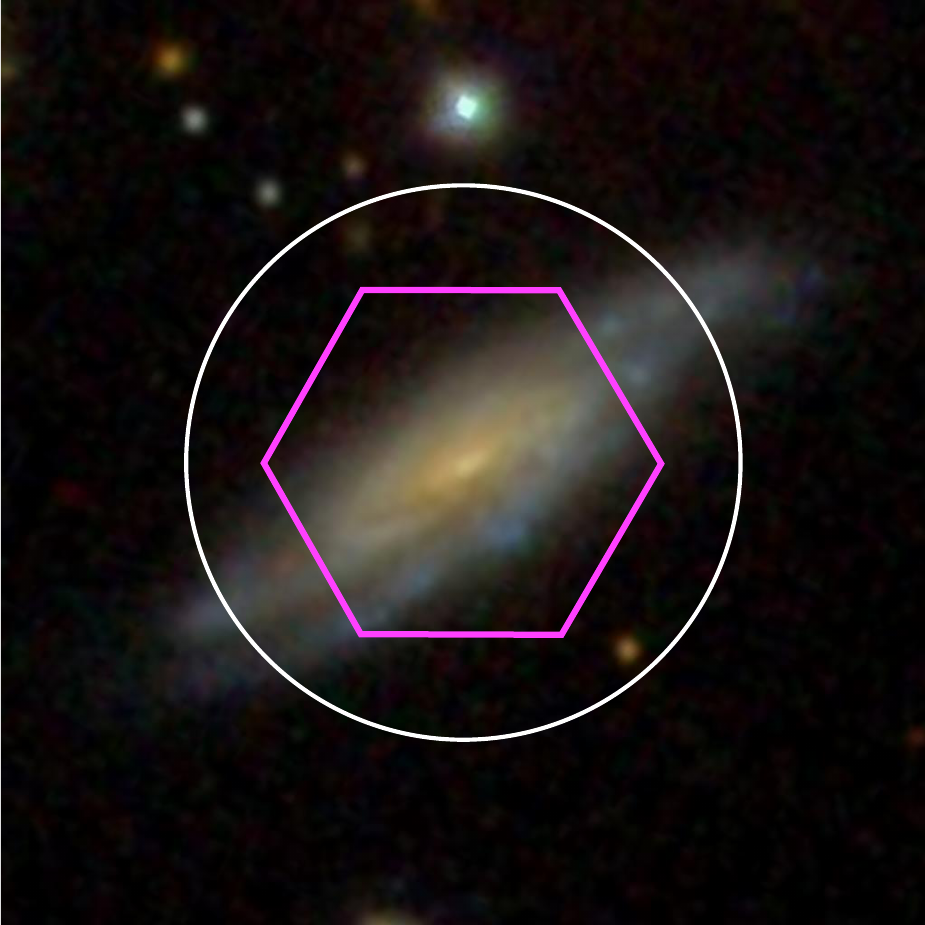}}
		\subfigure[]{\label{fig_L2}\includegraphics[width=0.32\textwidth]{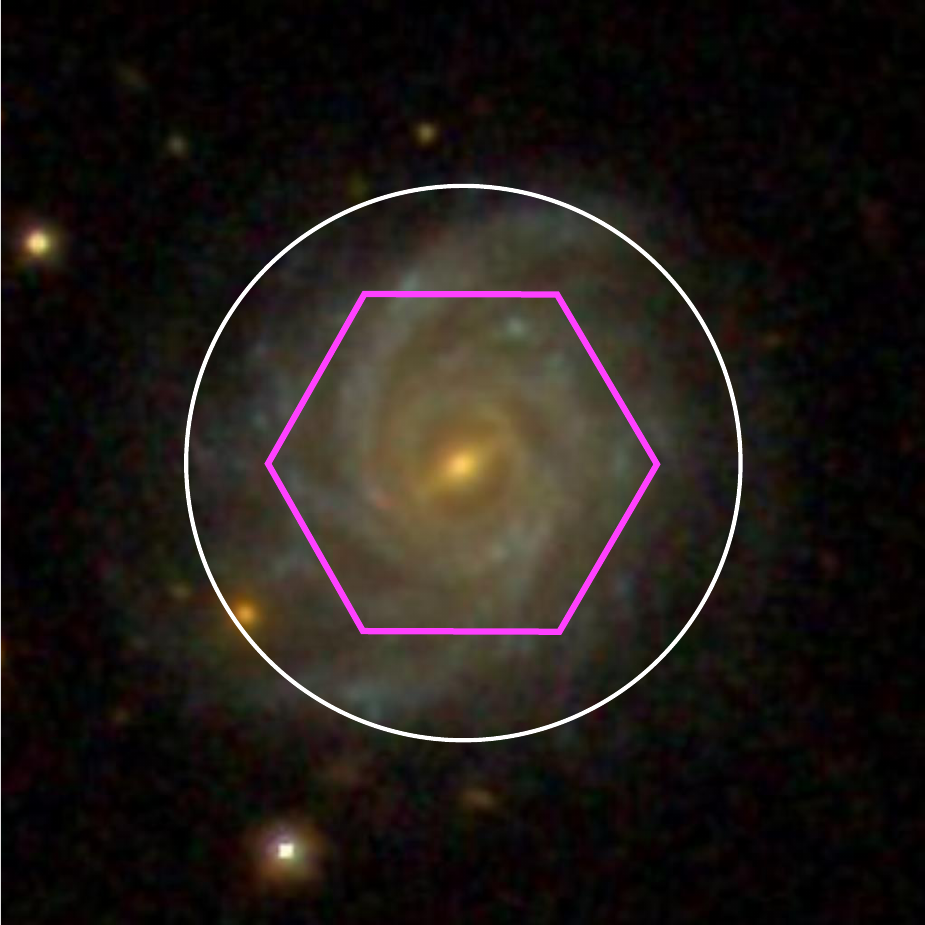}}
		\subfigure[]{\label{fig_L3}\includegraphics[width=0.32\textwidth]{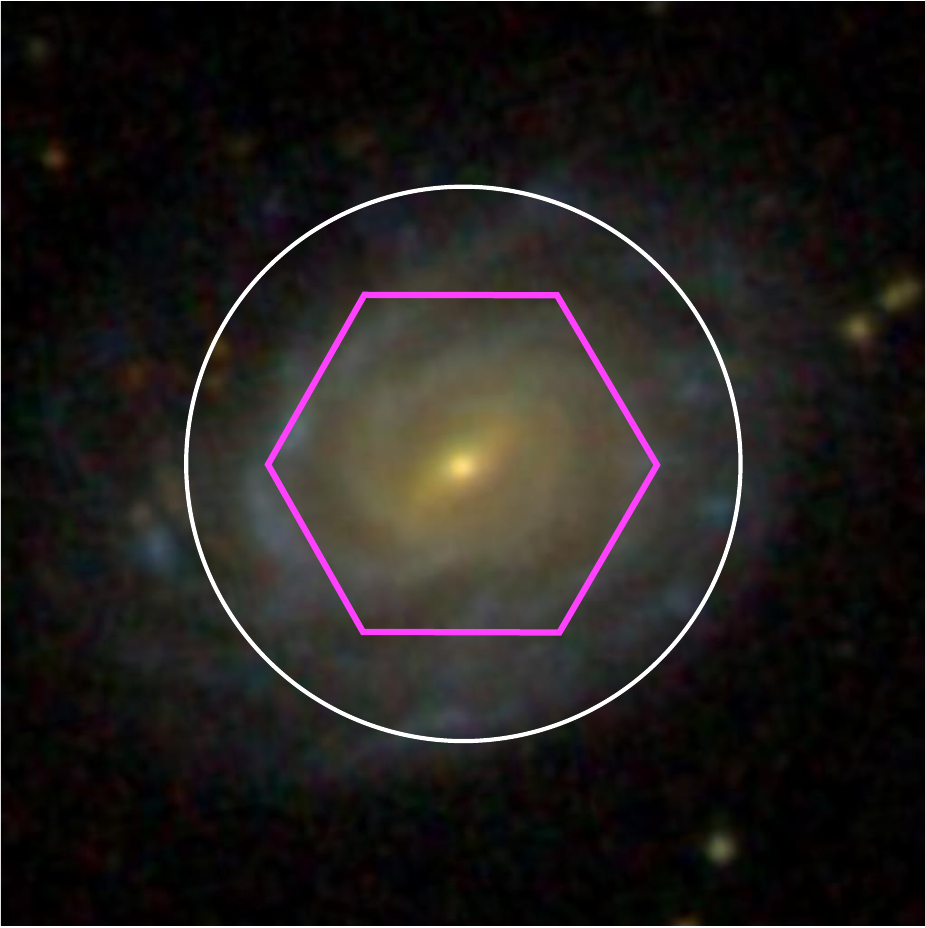}}
	\end{center}
	\caption{Varying galaxy sizes for galaxies with similar mass and redshift. All the galaxies in the figure have comparable \Mstar\ (log($M/M_{\sun}$) $\approx$ 10.5) and redshift ($\sim$ 0.02). However, the effective radii of the three galaxies in the upper row (relatively smooth, early-type disk galaxies) are 1.5 to 2 times smaller than those of the galaxies in the lower row (Milky Way-like late-type spirals). The white circles indicate an area with a diameter of 12 kpc for comparison purpose. The magenta hexagons show the MaNGA IFU coverage. These comparisons highlight the necessity of including galaxy size in the matching scheme to avoid comparing different galaxy populations.
	}
	\label{fig_match_size}
\end{figure*}

Finally, the importance of including galaxy size in the matching process is illustrated in Figure \ref{fig_match_size}. 
The six example galaxies shown in the figure have comparable 
\Mstar\ (log($M_\ast/M_{\sun}$) $\approx$ 10.5) and redshift ($\sim$ 0.02), but the sizes (effective radii) of the three galaxies in the upper row are 1.5 to 2 times smaller than those of the galaxies in the lower row. 
The smaller galaxies tend to be relatively smooth, early-type disk galaxies, while the larger galaxies resemble Milky Way-like late-type spirals, despite having similar \Mstar\ and redshift.
This variation in morphology highlight the necessity of including galaxy size in our matching scheme to ensure that we are comparing like with like and not conflating different galaxy populations.

\end{CJK}

\begin{thebibliography}{999}
\bibitem[Acharyya et al.(2020)]{Ach20} Acharyya, A., Krumholz, M.~R., Federrath, C., et al.\ 2020, \mnras, 495, 3819. doi:10.1093/mnras/staa1100
\bibitem[Aguado et al.(2018)]{Agu18} Aguado, D.~S., Ahumada, R., Almeida, A., et al.\ 2018, arXiv:1812.02759
\bibitem[Athanassoula(1992)]{Ath92} Athanassoula, E.\ 1992, \mnras, 259, 345. doi:10.1093/mnras/259.2.345
\bibitem[Athanassoula et al.(2016)]{Ath16} Athanassoula, E., Rodionov, S.~A., Peschken, N., et al.\ 2016, \apj, 821, 90. doi:10.3847/0004-637X/821/2/90
\bibitem[Barrera-Ballesteros et al.(2015)]{Bar15} Barrera-Ballesteros, J.~K., S{\'a}nchez, S.~F., Garc{\'\i}a-Lorenzo, B., et al.\ 2015, \aap, 579, A45. doi:10.1051/0004-6361/201425397
\bibitem[Barrera-Ballesteros et al.(2016)]{Bar16} Barrera-Ballesteros, J.~K., Heckman, T.~M., Zhu, G.~B., et al.\ 2016, \mnras, 463, 2513. doi:10.1093/mnras/stw1984
\bibitem[Bassini et al.(2024)]{Bas24} Bassini, L., Feldmann, R., Gensior, J., et al.\ 2024, \mnras, 532, L14. doi:10.1093/mnrasl/slae036
\bibitem[Belfiore et al.(2017)]{Bel17} Belfiore, F., Maiolino, R., Tremonti, C., et al.\ 2017, \mnras, 469, 151. doi:10.1093/mnras/stx789
\bibitem[Belfiore et al.(2019)]{Bel19} Belfiore, F., Westfall, K.~B., Schaefer, A., et al.\ 2019, \aj, 158, 160. doi:10.3847/1538-3881/ab3e4e
\bibitem[Boardman et al.(2023)]{Boa23} Boardman, N., Wild, V., Heckman, T., et al.\ 2023, \mnras, 520, 4301. doi:10.1093/mnras/stad277
\bibitem[Boardman et al.(2024)]{Boa24} Boardman, N., Wild, V., Rowlands, K., et al.\ 2024, \mnras, 527, 10788. doi:10.1093/mnras/stad3932
\bibitem[Bois et al.(2010)]{Boi10} Bois, M., Bournaud, F., Emsellem, E., et al.\ 2010, \mnras, 406, 2405. doi:10.1111/j.1365-2966.2010.16885.x
\bibitem[Bournaud et al.(2007)]{Bou07} Bournaud, F., Jog, C.~J., \& Combes, F.\ 2007, \aap, 476, 1179. doi:10.1051/0004-6361:20078010
\bibitem[Bundy et al.(2015)]{Bun15} Bundy, K., Bershady, M.~A., Law, D.~R., et al.\ 2015, \apj, 798, 7. doi:10.1088/0004-637X/798/1/7
\bibitem[Bustamante et al.(2018)]{Bus18} Bustamante, S., Sparre, M., Springel, V., et al.\ 2018, \mnras, 479, 3381. doi:10.1093/mnras/sty1692
\bibitem[Cavanagh \& Bekki(2020)]{Cav20} Cavanagh, M.~K. \& Bekki, K.\ 2020, \aap, 641, A77. doi:10.1051/0004-6361/202037963
\bibitem[Chang et al.(2022)]{Cha22} Chang, Y.-Y., Lin, L., Pan, H.-A., et al.\ 2022, \apj, 937, 97. doi:10.3847/1538-4357/ac8c27
\bibitem[Chen et al.(2023)]{Che23} Chen, Q.-H., Grasha, K., Battisti, A.~J., et al.\ 2023, \mnras, 519, 4801. doi:10.1093/mnras/stac3790
\bibitem[Cid Fernandes et al.(2013)]{Cid13} Cid Fernandes, R., P{\'e}rez, E., Garc{\'{\i}}a Benito, R., et al.\ 2013, \aap, 557, A86
\bibitem[Darg et al.(2010a)]{Dar10a} Darg, D.~W., Kaviraj, S., Lintott, C.~J., et al.\ 2010a, \mnras, 401, 1043
\bibitem[Darg et al.(2010b)]{Dar10b} Darg, D.~W., Kaviraj, S., Lintott, C.~J., et al.\ 2010b, \mnras, 401, 1552
\bibitem[Di Matteo et al.(2007)]{Dim07} Di Matteo, P., Combes, F., Melchior, A.-L., et al.\ 2007, \aap, 468, 61. doi:10.1051/0004-6361:20066959
\bibitem[Di Matteo et al.(2009)]{Dim09} Di Matteo, P., Pipino, A., Lehnert, M.~D., et al.\ 2009, \aap, 499, 427. doi:10.1051/0004-6361/200911715
\bibitem[Dom{\'\i}nguez S{\'a}nchez et al.(2022)]{Dom22} Dom{\'\i}nguez S{\'a}nchez, H., Margalef, B., Bernardi, M., et al.\ 2022, \mnras, 509, 4024. doi:10.1093/mnras/stab3089
\bibitem[Dopita et al.(2016)]{Dop16} Dopita, M.~A., Kewley, L.~J., Sutherland, R.~S., et al.\ 2016, \apss, 361, 61. doi:10.1007/s10509-016-2657-8
\bibitem[Drory et al.(2015)]{Dro15} Drory, N., MacDonald, N., Bershady, M.~A., et al.\ 2015, \aj, 149, 77 
\bibitem[Dutil \& Roy(1999)]{Dut99} Dutil, Y. \& Roy, J.-R.\ 1999, \apj, 516, 62. doi:10.1086/307100
\bibitem[Ellison et al.(2008)]{Ell08} Ellison, S.~L., Patton, D.~R., Simard, L., et al.\ 2008, \aj, 135, 1877. doi:10.1088/0004-6256/135/5/1877
\bibitem[Ellison et al.(2018)]{Ell18} Ellison, S.~L., S{\'a}nchez, S.~F., Ibarra-Medel, H., et al.\ 2018, \mnras, 474, 2039. doi:10.1093/mnras/stx2882
\bibitem[Emsellem et al.(2022)]{Ems22} Emsellem, E., Schinnerer, E., Santoro, F., et al.\ 2022, \aap, 659, A191. doi:10.1051/0004-6361/202141727
\bibitem[F{\"o}rster Schreiber et al.(2019)]{For19} F{\"o}rster Schreiber, N.~M., {\"U}bler, H., Davies, R.~L., et al.\ 2019, \apj, 875, 21. doi:10.3847/1538-4357/ab0ca2
\bibitem[Friedli et al.(1994)]{Fri94} Friedli, D., Benz, W., \& Kennicutt, R.\ 1994, \apjl, 430, L105. doi:10.1086/187449
\bibitem[Garay-Solis et al.(2023)]{Gar23} Garay-Solis, Y., Barrera-Ballesteros, J.~K., Colombo, D., et al.\ 2023, \apj, 952, 122. doi:10.3847/1538-4357/acd781
\bibitem[Garay-Solis et al.(2024)]{Gar24} Garay-Solis, Y., Barrera-Ballesteros, J.~K., Carigi, L., et al.\ 2024, \mnras, 533, 880. doi:10.1093/mnras/stae1876
\bibitem[Grasha et al.(2022)]{Gra22} Grasha, K., Chen, Q.~H., Battisti, A.~J., et al.\ 2022, \apj, 929, 118. doi:10.3847/1538-4357/ac5ab2
\bibitem[Hopkins et al.(2009)]{Hop09} Hopkins, P.~F., Cox, T.~J., Younger, J.~D., et al.\ 2009, \apj, 691, 1168. doi:10.1088/0004-637X/691/2/1168
\bibitem[Hwang et al.(2019)]{Hwa19} Hwang, H.-C., Barrera-Ballesteros, J.~K., Heckman, T.~M., et al.\ 2019, \apj, 872, 144. doi:10.3847/1538-4357/aaf7a3
\bibitem[Iono et al.(2004)]{Ion04} Iono, D., Yun, M.~S., \& Mihos, J.~C.\ 2004, \apj, 616, 199. doi:10.1086/424797
\bibitem[Karera et al.(2022)]{Kar22} Karera, P., Drissen, L., Martel, H., et al.\ 2022, \mnras, 514, 2769. doi:10.1093/mnras/stac1486
\bibitem[Kauffmann et al.(2003)]{Kau03} Kauffmann, G., Heckman, T.~M., White, S.~D.~M., et al.\ 2003, \mnras, 341, 33
\bibitem[Kewley et al.(2001)]{Kew01} Kewley, L.~J., Dopita, M.~A., Sutherland, R.~S., Heisler, C.~A., \& Trevena, J.\ 2001, \apj, 556, 121
\bibitem[Kewley \& Dopita(2002)]{Kew02} Kewley, L.~J. \& Dopita, M.~A.\ 2002, \apjs, 142, 35. doi:10.1086/341326
\bibitem[Kewley et al.(2006)]{Kew06} Kewley, L.~J., Geller, M.~J., \& Barton, E.~J.\ 2006, \aj, 131, 2004. doi:10.1086/500295
\bibitem[Kewley \& Ellison(2008)]{Kew08} Kewley, L.~J. \& Ellison, S.~L.\ 2008, \apj, 681, 1183. doi:10.1086/587500
\bibitem[Kewley et al.(2010)]{Kew10} Kewley, L.~J., Rupke, D., Zahid, H.~J., et al.\ 2010, \apjl, 721, L48. doi:10.1088/2041-8205/721/1/L48
\bibitem[Kewley et al.(2019)]{Kew19} Kewley, L.~J., Nicholls, D.~C., \& Sutherland, R.~S.\ 2019, \araa, 57, 511. doi:10.1146/annurev-astro-081817-051832
\bibitem[Kim et al.(2021)]{Kim21} Kim, E., Hwang, H.~S., Jeong, W.-S., et al.\ 2021, \mnras, 507, 3113. doi:10.1093/mnras/stab2090
\bibitem[Kumari et al.(2019)]{Kum19} Kumari, N., Maiolino, R., Belfiore, F., et al.\ 2019, \mnras, 485, 367. doi:10.1093/mnras/stz366
Tinsley, B.~M.\ 1978, \apj, 219, 46 
\bibitem[Larson(1976)]{Lar76} Larson, R.~B.\ 1976, \mnras, 176, 31. doi:10.1093/mnras/176.1.31
\bibitem[Law et al.(2015)]{Law15} Law, D.~R., Yan, R., Bershady, M.~A., et al.\ 2015, \aj, 150, 19 
\bibitem[Law et al.(2016)]{Law16} Law, D.~R., Cherinka, B., Yan, R., et al.\ 2016, \aj, 152, 83
\bibitem[Lin et al.(2004)]{Lin04} Lin, L., Koo, D.~C., Willmer, C.~N.~A., et al.\ 2004, \apjl, 617, L9
\bibitem[Lin et al.(2008)]{Lin08} Lin, L., Patton, D.~R., Koo, D.~C., et al.\ 2008, \apj, 681, 232. doi:10.1086/587928
\bibitem[Lotz et al.(2008a)]{Loz08a} Lotz, J.~M., Jonsson, P., Cox, T.~J., et al.\ 2008a, \mnras, 391, 1137. doi:10.1111/j.1365-2966.2008.14004.x
\bibitem[Lotz et al.(2008b)]{Loz08b} Lotz, J.~M., Davis, M., Faber, S.~M., et al.\ 2008b, \apj, 672, 177. doi:10.1086/523659
\bibitem[Maiolino \& Mannucci(2019)]{Mao19} Maiolino, R. \& Mannucci, F.\ 2019, \aapr, 27, 3. doi:10.1007/s00159-018-0112-2
\bibitem[Marino et al.(2013)]{Mar13} Marino, R.~A., Rosales-Ortega, F.~F., S{\'a}nchez, S.~F., et al.\ 2013, \aap, 559, A114. doi:10.1051/0004-6361/201321956
\bibitem[Martinet \& Friedli(1997)]{Mar97} Martinet, L. \& Friedli, D.\ 1997, \aap, 323, 363. doi:10.48550/arXiv.astro-ph/9701091
\bibitem[Mast et al.(2014)]{Mas14} Mast, D., Rosales-Ortega, F.~F., S{\'a}nchez, S.~F., et al.\ 2014, \aap, 561, A129. doi:10.1051/0004-6361/201321789
\bibitem[Mingozzi et al.(2020)]{Min20} Mingozzi, M., Belfiore, F., Cresci, G., et al.\ 2020, \aap, 636, A42. doi:10.1051/0004-6361/201937203
\bibitem[Moetazedian et al.(2017)]{Moe17} Moetazedian, R., Polyachenko, E.~V., Berczik, P., et al.\ 2017, \aap, 604, A75. doi:10.1051/0004-6361/201630024
\bibitem[Montuori et al.(2010)]{Mon10} Montuori, M., Di Matteo, P., Lehnert, M.~D., et al.\ 2010, \aap, 518, A56. doi:10.1051/0004-6361/201014304
\bibitem[Moreno et al.(2021)]{Mor21} Moreno, J., Torrey, P., Ellison, S.~L., et al.\ 2021, \mnras, 503, 3113. doi:10.1093/mnras/staa2952
\bibitem[Nevin et al.(2023)]{Nev23} Nevin, R., Blecha, L., Comerford, J., et al.\ 2023, \mnras, 522, 1. doi:10.1093/mnras/stad911
\bibitem[Newman et al.(2012)]{New12} Newman, S.~F., Genzel, R., F{\"o}rster-Schreiber, N.~M., et al.\ 2012, \apj, 761, 43. doi:10.1088/0004-637X/761/1/43
\bibitem[Pan et al.(2019)]{Pan19} Pan, H.-A., Lin, L., Hsieh, B.-C., et al.\ 2019, \apj, 881, 119. doi:10.3847/1538-4357/ab311c
\bibitem[Patton et al.(2002)]{Pat02} Patton, D.~R., Pritchet, C.~J., Carlberg, R.~G., et al.\ 2002, \apj, 565, 208  
\bibitem[Peebles(1969)]{Pee69} Peebles, P.~J.~E.\ 1969, \apj, 155, 393. doi:10.1086/149876
\bibitem[Perez et al.(2011)]{Per11} Perez, J., Michel-Dansac, L., \& Tissera, P.~B.\ 2011, \mnras, 417, 580. doi:10.1111/j.1365-2966.2011.19300.x
\bibitem[P{\'e}rez-Montero et al.(2016)]{Per16} P{\'e}rez-Montero, E., Garc{\'\i}a-Benito, R., V{\'\i}lchez, J.~M., et al.\ 2016, \aap, 595, A62. doi:10.1051/0004-6361/201628601
\bibitem[Pettini \& Pagel(2004)]{Pet04} Pettini, M. \& Pagel, B.~E.~J.\ 2004, \mnras, 348, L59. doi:10.1111/j.1365-2966.2004.07591.x
\bibitem[Poetrodjojo et al.(2018)]{Poe18} Poetrodjojo, H., Groves, B., Kewley, L.~J., et al.\ 2018, \mnras, 479, 5235. doi:10.1093/mnras/sty1782
\bibitem[Porter et al.(2022)]{Por22} Porter, L.~E., Orr, M.~E., Burkhart, B., et al.\ 2022, \mnras, 515, 3555. doi:10.1093/mnras/stac1958
\bibitem[Robertson et al.(2006)]{Rob06} Robertson, B., Bullock, J.~S., Cox, T.~J., et al.\ 2006, \apj, 645, 986. doi:10.1086/504412
\bibitem[Rosales-Ortega et al.(2012)]{Ros12} Rosales-Ortega, F.~F., S{\'a}nchez, S.~F., Iglesias-P{\'a}ramo, J., et al.\ 2012, \apjl, 756, L31. doi:10.1088/2041-8205/756/2/L31
\bibitem[Rupke et al.(2010)]{Rup10} Rupke, D.~S.~N., Kewley, L.~J., \& Barnes, J.~E.\ 2010, \apjl, 710, L156. doi:10.1088/2041-8205/710/2/L156
\bibitem[S{\'a}nchez et al.(2014)]{San14} S{\'a}nchez, S.~F., Rosales-Ortega, F.~F., Iglesias-P{\'a}ramo, J., et al.\ 2014, \aap, 563, A49. doi:10.1051/0004-6361/201322343
\bibitem[S{\'a}nchez et al.(2016a)]{San16a} S{\'a}nchez, S.~F., P{\'e}rez, E., S{\'a}nchez-Bl{\'a}zquez, P., et al.\ 2016a, \rmxaa, 52, 21 
\bibitem[S{\'a}nchez et al.(2016b)]{San16b} S{\'a}nchez, S.~F., P{\'e}rez, E., S{\'a}nchez-Bl{\'a}zquez, P., et al.\ 2016b, \rmxaa, 52, 171 
\bibitem[S{\'a}nchez et al.(2018)]{San18} S{\'a}nchez, S.~F., Avila-Reese, V., Hernandez-Toledo, H., et al.\ 2018, \rmxaa, 54, 217
\bibitem[S{\'a}nchez(2020)]{San20} S{\'a}nchez, S.~F.\ 2020, \araa, 58, 99. doi:10.1146/annurev-astro-012120-013326
\bibitem[Scudder et al.(2012)]{Scu12} Scudder, J.~M., Ellison, S.~L., Torrey, P., et al.\ 2012, \mnras, 426, 549. doi:10.1111/j.1365-2966.2012.21749.x
\bibitem[Scudder et al.(2021)]{Scu21} Scudder, J.~M., Ellison, S.~L., El Meddah El Idrissi, L., et al.\ 2021, \mnras, 507, 2468. doi:10.1093/mnras/stab2339
\bibitem[Sharda et al.(2021a)]{Sha21a} Sharda, P., Krumholz, M.~R., Wisnioski, E., et al.\ 2021a, \mnras, 502, 5935. doi:10.1093/mnras/stab252
\bibitem[Sharda et al.(2021b)]{Sha21b} Sharda, P., Krumholz, M.~R., Wisnioski, E., et al.\ 2021b, \mnras, 504, 53. doi:10.1093/mnras/stab868
\bibitem[Sharda et al.(2024)]{Sha24} Sharda, P., Ginzburg, O., Krumholz, M.~R., et al.\ 2024, \mnras, 528, 2232. doi:10.1093/mnras/stae088
\bibitem[Sinha \& Holley-Bockelmann(2012)]{Sin12} Sinha, M. \& Holley-Bockelmann, K.\ 2012, \apj, 751, 17. doi:10.1088/0004-637X/751/1/17
\bibitem[Tapia-Contreras et al.(2025)]{Tap25} Tapia-Contreras, B., Tissera, P.~B., Sillero, E., et al.\ 2025, arXiv:2502.02080. doi:10.48550/arXiv.2502.02080
\bibitem[Teimoorinia et al.(2021)]{Tei21} Teimoorinia, H., Jalilkhany, M., Scudder, J.~M., et al.\ 2021, \mnras, 503, 1082. doi:10.1093/mnras/stab466
\bibitem[Tissera et al.(2019)]{Tis19} Tissera, P.~B., Rosas-Guevara, Y., Bower, R.~G., et al.\ 2019, \mnras, 482, 2208. doi:10.1093/mnras/sty2817
\bibitem[Tissera et al.(2022)]{Tis22} Tissera, P.~B., Rosas-Guevara, Y., Sillero, E., et al.\ 2022, \mnras, 511, 1667. doi:10.1093/mnras/stab3644
\bibitem[Tissera et al.(2025)]{Tis25} Tissera, P., Bignone, L., Gonzalez-Jara, J., et al.\ 2025, arXiv:2501.05978. doi:10.48550/arXiv.2501.05978
\bibitem[Thorp et al.(2019)]{Tho19} Thorp, M.~D., Ellison, S.~L., Simard, L., et al.\ 2019, \mnras, 482, L55. doi:10.1093/mnrasl/sly185
\bibitem[Toomre(1977)]{Too77} Toomre, A.\ 1977, Evolution of Galaxies and Stellar Populations, 401
\bibitem[Torrey et al.(2012)]{Tor12} Torrey, P., Cox, T.~J., Kewley, L., et al.\ 2012, \apj, 746, 108. doi:10.1088/0004-637X/746/1/108
\bibitem[Torrey et al.(2019)]{Tor19} Torrey, P., Vogelsberger, M., Marinacci, F., et al.\ 2019, \mnras, 484, 5587. doi:10.1093/mnras/stz243
\bibitem[Tremonti et al.(2004)]{Tre04} Tremonti, C.~A., Heckman, T.~M., Kauffmann, G., et al.\ 2004, \apj, 613, 898. doi:10.1086/423264
\bibitem[Vale Asari et al.(2019)]{Val19} Vale Asari, N., Couto, G.~S., Cid Fernandes, R., et al.\ 2019, \mnras, 489, 4721. doi:10.1093/mnras/stz2470
\bibitem[Wake et al.(2017)]{Wak17} Wake, D.~A., Bundy, K., Diamond-Stanic, A.~M., et al.\ 2017, \aj, 154, 86
\bibitem[Wang \& Lilly(2021)]{Wan21} Wang, E. \& Lilly, S.~J.\ 2021, \apj, 910, 137. doi:10.3847/1538-4357/abe413
\bibitem[Westfall et al.(2019)]{Wes19} Westfall, K.~B., Cappellari, M., Bershady, M.~A., et al.\ 2019, arXiv:1901.00856 
\bibitem[Zinchenko et al.(2019)]{Zin19} Zinchenko, I.~A., Just, A., Pilyugin, L.~S., et al.\ 2019, \aap, 623, A7. doi:10.1051/0004-6361/201834364
\bibitem[Zurita et al.(2021)]{Zur21} Zurita, A., Florido, E., Bresolin, F., et al.\ 2021, \mnras, 500, 2380. doi:10.1093/mnras/staa2208
\end{thebibliography}
\end{document}